\documentclass[journal=jctcce,manuscript=article,layout=twocolumn]{achemso}
\setkeys{acs}{etalmode=truncate,maxauthors=4}
\setkeys{acs}{articletitle=true}

\usepackage{chemformula} % Formula subscripts using \ch{}
\usepackage[T1]{fontenc} % Use modern font encodings
\usepackage{amsmath,amsfonts}
\usepackage{physics}
\usepackage{comment}
\usepackage{booktabs}
\usepackage{multirow}
\usepackage{textcomp}  % \textdegree symbol
\mciteErrorOnUnknownfalse
\usepackage{caption}
\usepackage{subcaption}
\usepackage{widetext}
\usepackage[version=4]{mhchem}
\usepackage{qcircuit}
\usepackage[draft]{revision}
\usepackage{letltxmacro}
\LetLtxMacro{\originaleqref}{\eqref}
\renewcommand{\eqref}{Eq.\,\originaleqref}

\mathchardef\mhyphen="2D
\newcommand{\diff}{\,\mathrm{d}}

\newcommand*\Hop{\widehat{H}}
\newcommand*\Uop{\widehat{U}}

\newcommand*\Qop{\widehat{Q}}

\newcommand*\aop{\widehat{a}}

\newcommand*\Norb{N_{\text{orb}}}

\newcommand*\bbC{\mathbb{C}}
\newcommand*\bbR{\mathbb{R}}
\newcommand*\bbP{\mathbb{P}}

\newcommand*\eg{{\it e.g.}}
\newcommand*\ie{{\it i.e.}}

\author{Joel Bierman}
\affiliation[Duke]{Department of Physics, Duke University}
\author{Yingzhou Li}
\affiliation[Fudan]{School of Mathematical Sciences, Fudan University}
\email{yingzhouli@fudan.edu.cn}
\author{Jianfeng Lu}
\email{jianfeng@math.duke.edu}
\affiliation[Duke]{Department of Mathematics, Duke University}
\alsoaffiliation{Department of Physics, Duke University}
\alsoaffiliation{Department of Chemistry, Duke University}

\title[Quantum Excited States]{Quantum Orbital Minimization Method
for Excited States Calculation on Quantum Computer}

\abbreviations{}
\keywords{full configuration interaction, excited state energy;
eigenvalue}

\begin{document}
% \begin{tocentry}
% 
%     Some journals require a graphical entry for the Table of Contents.
%     This should be laid out ``print ready'' so that the sizing of the
%     text is correct.
% 
%     Inside the \texttt{tocentry} environment, the font used is Helvetica
%     8\,pt, as required by \emph{Journal of the American Chemical
%     Society}.
% 
%     The surrounding frame is 9\,cm by 3.5\,cm, which is the maximum
%     permitted for  \emph{Journal of the American Chemical Society}
%     graphical table of content entries. The box will not resize if the
%     content is too big: instead it will overflow the edge of the box.
% 
%     This box and the associated title will always be printed on a
%     separate page at the end of the document.
% 
% \end{tocentry}

\begin{abstract}

We propose a quantum-classical hybrid variational algorithm, the quantum
orbital minimization method (qOMM), for obtaining the ground state and
low-lying excited states of a Hermitian operator. Given parameterized
ansatz circuits representing eigenstates, qOMM implements quantum circuits
to represent the objective function in the orbital minimization method
and adopts a classical optimizer to minimize the objective function with
respect to the parameters in ansatz circuits. The objective function has
orthogonality constraint implicitly embedded, which allows qOMM to apply
a different ansatz circuit to each input reference state. We carry out
numerical simulations that seek to find excited states of \ch{H_2}, \ch{LiH}, and a toy model consisting of 4 hydrogen atoms arranged in a square lattice in the STO-3G basis with UCCSD
ansatz circuits. Comparing the numerical results with existing excited states
methods, qOMM is less prone to getting stuck in local minima and can achieve
convergence with more shallow ansatz circuits.

\end{abstract}

\section{Introduction}
\label{sec:intro}

The development of quantum computing has boomed in recent years. The
supremacy of quantum computing is demonstrated by several
groups~\cite{Arute2019, Zhong2020}. One of the most promising applications
for noisy intermediate-scale quantum~(NISQ) devices is the variational
quantum eigensolver~(VQE) under the full configuration interaction~(FCI)
framework~\cite{Blunt2015, Holmes2017, Schriber2017, Li2019c, Wang2019,
Li2020a}. FCI is a quantum chemistry method which discretizes the
time-independent many-body Schr{\"o}dinger equation numerically exactly.
The ground state and low-lying excited states are calculated via solving a
standard eigenvalue problem, while the problem dimension scales
exponentially as the number of electrons in the system. The FCI framework
fits naturally into quantum computing. Quantum algorithms for ground state
eigensolvers under FCI framework have been extensively developed in past
decades~\cite{Abrams1999, Aspuru-Guzik2005, Peruzzo2014, Bauer2020,
McArdle2020}. Among them, VQE is a quantum-classical hybrid method and has
the shortest circuit depth, which makes it the most widely applied ground
state quantum eigensolver on NISQ devices. In short, VQE adopts a
parameterized circuit as the variational ansatz $\ket{\phi_\theta}$ for the
ground state. Then the energy function $\expval*{\Hop}{\phi_\theta}$ is
minimized with respect to the parameter $\theta$, where $\Hop$ is the
Hamiltonian operator.

In addition to the ground state, excited states play an important role in
connecting computational results and experimental observations in quantum
chemistry. In this paper, we propose a hybrid quantum-classical algorithm,
named quantum orbital minimization method~(qOMM), to compute the excited
state energies as well as the corresponding states under FCI framework.

Existing approaches for excited states on quantum computers could be
grouped into two categories: the excited state energy computation and the
excited state vector computation.

Excited state energy computation on quantum computer evaluates the excited
state energy without explicitly constructing the state vector. Two
methods, quantum subspace expansion~(QSE)~\cite{McClean2017a, Colless2018,
Ganzhorn2019} and quantum equation of motion~(qEoM)~\cite{Ollitrault2020},
are representatives of this category. Both of them are based on
perturbation theory starting from the ground state. They first construct
the ground state $\ket{\phi_{GS}}$ via VQE. Then they evaluate the
expansion values on single excited states from the ground state
$\ket{\phi_{GS}}$, \ie,
\begin{equation*}
    A^{p_2 q_2}_{p_1 q_1} = \ev**{\aop_{q_2}^\dagger
    \aop_{p_2} \Hop \aop_{p_1}^\dagger \aop_{q_1}}{\phi_{GS}}
\end{equation*}
and
\begin{equation*}
    B^{p_2 q_2}_{p_1 q_1} = \ev**{\aop_{q_2}^\dagger \aop_{p_2}
    \aop_{p_1}^\dagger \aop_{q_1}}{\phi_{GS}}
\end{equation*}
for $\aop_{p}^\dagger$ and $\aop_{p}$ being creation and annihilation
operators. Once these expansion values are available, QSE solves a
generalized eigenvalue problem of matrix pencil $(A,B)$ and obtains the
refined ground state energy and low-lying excited state energies on the
perturbed space of $\ket{\phi_{GS}}$. qEoM adopts these expansion values
into the equation of motion expression and obtains the low-lying excited
state energies. Double excitations and higher order excitations from the
ground state could be further evaluated and included to improve the
accuracy of QSE and qEoM. However, including more excitations leads to a
much higher computational cost.

Excited state vector computations include a wide spectrum of methods.
Folded spectrum method~\cite{McClean2016} folds the spectrum of a
Hamiltonian operator via a shift-and-square operation, \ie, $\Hop
\rightarrow (\Hop - \lambda I)^2$, where $\lambda$ is a chosen number
close to the desired excited state energy. Then $(\Hop - \lambda I)^2$ is
used as the operator in VQE method, and an excited state with the energy
closest to $\lambda$ is obtained via the optimization procedure in VQE.
Witness-assisted variational eigenspectra solver~\cite{Santagati2017,
Santagati2018} adopts an entropy term as the regularizer in the objective
function to keep the minimized excited state close to its initial state,
where the initial state is set to be an excitation from the ground state.
Quantum deflation method~\cite{Higgott2019, Jones2019} adopts the
overlapping between the target state and the ground state as a penalty
term in the objective function to keep the target state as orthogonal as
possible to the ground state. The extension to more than one excited state
in the quantum deflation method is straightforward. All these methods
above have their classical algorithm counterparts. On the other hand, the
following two methods use a unique property of quantum computing, \ie, all
quantum operators are unitary and orthogonal states after the actions of
quantum operators are still orthogonal to each other. Subspace search
VQE~(SSVQE)~\cite{Nakanishi2019} and multistate contracted
VQE~(MCVQE)~\cite{Parrish2019} apply the same basic ansatz. They first
prepare a sequence of non-parameterized mutually orthogonal states,
$\ket{\phi_1}, \ket{\phi_2}, \dots, \ket{\phi_K}$ for $K$ excited states.
Then a parameterized circuit $\Uop_\theta$ is applied to these states, as
$\Uop_\theta$ is unitary (implemented on a quantum computer), $\Uop_\theta
\ket{\phi_1}, \dots, \Uop_\theta \ket{\phi_K}$ are mutually orthogonal.
Hence we could minimize the objective function
\begin{equation*}
    \sum_k \ev**{\Uop^\dagger_\theta \Hop \Uop_\theta}{\phi_k}
\end{equation*}
to obtain approximated ground state and excited states. The
energies associated with $\Uop_\theta \ket{\phi_1}, \dots, \Uop_\theta
\ket{\phi_K}$ are usually not ordered. SSVQE further introduces a min-max
problem to obtain a specific excited state. Also, in the SSVQE paper,
a weighted objective function is employed to preserve the ordering, which will
be the version we investigate numerically in our paper below. The MCVQE
approach enriches the ansatz by introducing a unitary matrix applied across
states in a classical post-processing step, \ie, $\ket{\psi_k} =
\Uop_\theta \sum_{\ell} \ket{\phi_\ell} V_{\ell k}$ for $V_{\ell k}$ being
the $(\ell,k)$-th entry of a unitary matrix $V$. Hence a standard eigenvalue
problem is solved as the post-processing in MCVQE to obtain $V$.

In this paper, we propose to obtain the excited state vector using the
objective function in the orbital minimization
method~(OMM)~\cite{Ordejon1993, Mauri1993},
\begin{equation} \label{eq:omm-obj}
    \begin{split}
        f(\ket{\psi_1}, \dots, \ket{\psi_K}) =& 2 \sum_{i=1}^K
        \ev**{\Hop}{\psi_i} \\
        &- \sum_{i,j=1}^K \braket*{\psi_i}{\psi_j}
        \mel**{\psi_j}{\Hop}{\psi_i}.
    \end{split}
\end{equation}
We refer our method as quantum orbital minimization method~(qOMM) throughout
this paper. An important property of \eqref{eq:omm-obj} is that, even
though the orthogonality conditions among $\ket{\Psi_i}$s are not explicitly
enforced, the minimizer of \eqref{eq:omm-obj} consists of mutually orthogonal
state vectors. For most other methods, the orthogonality condition is
explicitly proposed as a constraint in the optimization problem. The implicit
orthogonality property allows us to adopt a different ansatz (with regards
to either the numerical parameter values, the structure of the ansatz, or
both) for each excited state. In this paper, the variational
ansatz class that we use is of the form,
\begin{equation}
    \ket{\widetilde{\psi}_k} = \sum_{\ell=1}^K \Uop_{\theta_\ell}
    \ket{\phi_\ell} V_{\ell k},
\end{equation}
where $\ket{\phi_1}, \dots, \ket{\phi_K}$ are initial states, $\Uop_{\theta_1},
\dots, \Uop_{\theta_K}$ are parameterized circuits with $\theta_1, \dots,
\theta_K$ being the parameters, $V_{\ell k}$ is the $(\ell, k)$-th entry of an
invertible matrix $V$ that is applied during a classical post-processing
step. Such an expression is more general than those used in other excited state vector methods, which typically either apply the same circuit $\Uop_{\theta}$
to all input states or do not apply the classical post-processing matrix $V$.
Another feature of \eqref{eq:omm-obj} is that the objective function
does not include any hyper-parameter (unlike, e.g., the deflation approach),
which makes the method easy to use in practice without parameter tuning. Furthermore, this approach shares the same advantages over QSE and qEoM as SSVQE and MCVQE. Namely, the excited states in qOMM are encoded into the objective function of an optimization problem and not as linear combinations of single and double excitations above the ground state in a post-processing step. This allows the method to 1. compute excited states for which triple and higher excitation terms have a non-negligible contribution without calculating higher order reduced density matrices and 2. treat the ground and excited states on the same footing.

Notice that the objective function \eqref{eq:omm-obj} includes the
evaluation of the state inner product with respect to the Hamiltonian
operator and the state inner product, \ie, $\mel*{\phi_i}{\Hop}{\psi_j}$
and $\braket*{\psi_i}{\psi_j}$, respectively. In other existing excited
state methods, neither of these two are evaluated directly.\footnote{In
deflation method~\cite{Higgott2019, Jones2019}, the absolute value of the
inner product, $\abs*{\braket*{\psi_i}{\psi_j}}$, is evaluated, while we need
the actual value of the inner product without taking the absolute value.}
Given our variational ansatz, we construct quantum circuits to evaluate both
inner products, which are essentially inspired by the Hadamard test. When
the evaluation of \eqref{eq:omm-obj} is available on a quantum computer,
we can adopt gradient-free optimizers to minimize the objective function
with respect to parameters, $\theta_1, \dots, \theta_K$.

Finally, we include a sequence of numerical results to demonstrate the
efficiency and robustness of the proposed method. We test the following
chemistry systems using the proposed method on quantum simulators: \ch{H_2}
and \ch{LiH} at their equilibrium configurations, \ch{H_2} at a stretched bond
distance of twice the equilibrium distance, and a toy model consisting of
4 hydrogen atoms arranged in a square at near-equilibrium distances and a stretched bond distance configuration. We observe that qOMM has two major
benefits compared to algorithms such as weighted-SSVQE (In this paper, when
we refer to SSVQE, we implicitly mean the weighted version of SSVQE that uses
one optimization step for finding multiple eigenvalues simultaneously.) that
enforce the orthogonality of the input states at every optimization step. 1)
Enforcing the mutual orthogonality of the input states through the overlap
terms allows for greater flexibility in the choice of ansatz applied to each
input state, as well as greater flexibility in the choice of input states. The
circuit applied to each input state must be identical in the SSVQE framework
in order to enforce orthogonality, whereas qOMM allows us to apply a different
ansatz circuit to each input state. 2) The tendency to get stuck in local
minima, while not eliminated entirely, is diminished considerably. The first
benefit is important because choosing a suitably expressive ansatz for excited
states is often more difficult than the ground state VQE problem. In practice,
this means that one is often able to use a more shallow and less expressive
ansatz circuit; a feature that is crucial for applications on NISQ devices.

The rest of the paper is organized as follows. Section~\ref{sec:innerprod}
illustrates the circuits we use to evaluate the inner products. The
overall method is discussed in Section~\ref{sec:method}. In
Section~\ref{sec:num}, we apply the method to various chemistry molecules
to demonstrate the efficiency of the proposed method. Finally,
Section~\ref{sec:conclusion} concludes the paper and discusses future
work.

\section{Inner Product Circuits}
\label{sec:innerprod}

In the objective function of qOMM, the inner product of two states and
the inner product of two states with respect to an operator are complex
numbers. Both the real and imaginary parts are explicitly needed to construct
the objective function. If two states are identical, then the inner products
can be efficiently evaluated by the well known Hadamard test. We propose quantum circuits for the
evaluation of the inner products for the case in which they are different
in this section.

To simplify the notations, we consider evaluating $\braket*{\psi}{\phi}$
and $\mel*{\psi}{\Hop}{\phi}$, where $\ket{\psi}$ and $\ket{\phi}$ are
given by a unitary circuit acting on $\ket{0}$, \ie, $\ket{\psi} =
\Uop_{\theta_\psi} \ket{0}$ and $\ket{\phi} = \Uop_{\theta_\phi} \ket{0}$.
Given the variational ansatz, the inner product $\braket*{\psi}{\phi} =
\ev**{\Uop_{\theta_\psi}^\dagger \Uop_{\theta_\phi}}{0}$ can be viewed as
the inner product of $\ket{0}$ with respect to the operator $\widehat{O} =
\Uop_{\theta_\psi}^\dagger \Uop_{\theta_\phi}$. In principle, the
transpose of the ansatz circuit can be constructed. After the ansatz
circuit is compiled into basic gates, the transpose of the circuit is the
composition of the transposed basic gates in the reverse ordering. Hence
$\braket*{\psi}{\phi}$ could be then evaluated using the Hadamard test,
which requires the controlled ansatz circuit and the controlled transpose
of the ansatz circuit. Instead, we adopt an idea inspired by the Hadamard
test, which can be applied to evaluate $\mel*{\psi}{\Hop}{\phi}$ without
controlled transpose of the ansatz circuit. In the following, we discuss
such two quantum circuits evaluating the inner products in detail.

\vspace{.1em} 

\subsection{$\braket*{\psi}{\phi}$ evaluation}

For complex-valued states $\ket{\psi}$ and $\ket{\phi}$, the inner product
$\braket*{\psi}{\phi}$ is also a complex number. The evaluation of
$\braket*{\psi}{\phi}$ is divided into the evaluations of the real and
imaginary parts. We will introduce the circuit for evaluation of the real
part of $\braket*{\psi}{\phi}$ in detail, and the circuit for the
imaginary part could be constructed analogously.

\begin{figure}[htp]
    \centering
    \includegraphics[height=4cm]{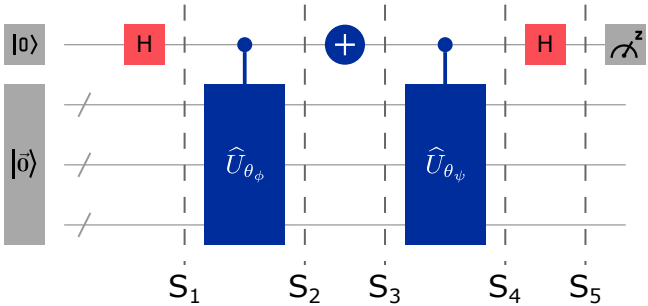}
    \caption{Inner product circuit.}
    \label{fig:inner_prod}
\end{figure}

Figure~\ref{fig:inner_prod} illustrates the precise quantum circuit used
to evaluate the real part of $\braket*{\psi}{\phi}$. Here we list
expressions at all five stages as shown in the figure:

\begin{widetext}
\begin{align}
    \Big(H \otimes I \Big) \ket{0} \ket*{\vec{0}}
    = &
    \frac{1}{\sqrt{2}} \Big(\ket{0} + \ket{1}\Big)\ket*{\vec{0}},
    \tag{S\textsubscript{1}}\\
    C\mhyphen\Uop_{\theta_\phi}
    \frac{1}{\sqrt{2}}
    \Big(\ket{0} + \ket{1}\Big)\ket*{\vec{0}}
    = &
    \frac{1}{\sqrt{2}} \Big(\ket{0}\ket*{\vec{0}} + \ket{1}\ket{\phi}\Big),
    \tag{S\textsubscript{2}} \label{eq:S2}\\
    \Big(X \otimes I\Big)
    \frac{1}{\sqrt{2}} \Big(\ket{0}\ket*{\vec{0}} + \ket{1}\ket{\phi}\Big)
    = &
    \frac{1}{\sqrt{2}} \Big(\ket{1}\ket*{\vec{0}} + \ket{0}\ket{\phi}\Big),
    \tag{S\textsubscript{3}}\\
    C\mhyphen\Uop_{\theta_\psi}
    \frac{1}{\sqrt{2}}
    \Big(\ket{1}\ket*{\vec{0}} + \ket{0}\ket{\phi}\Big)
    = &
    \frac{1}{\sqrt{2}}\Big( \ket{1}\ket{\psi} + \ket{0}\ket{\phi} \Big),
    \tag{S\textsubscript{4}}\\
    \Big(H \otimes I\Big)
    \frac{1}{\sqrt{2}}\Big( \ket{1}\ket{\psi} + \ket{0}\ket{\phi} \Big)
    = &
    \frac{1}{2} \Big( \ket{0} \big(\ket{\phi} + \ket{\psi}\big)
    + \ket{1} \big( \ket{\phi} - \ket{\psi} \big) \Big),
    \tag{S\textsubscript{5}}
\end{align}
\end{widetext}

\noindent
where $H$ is the Hadamard gate, $X$ is the Pauli-X gate, $I$ is the
identity mapping, $C\mhyphen\Uop$ is the controlled $\Uop$ gate.
Conducting the measurement on the ancilla qubit, we have the probability
measuring zero being,
\begin{equation} \label{eq:measure-real}
    \begin{split}
        \bbP(\text{measurement} & =0) \\
        & = \frac{1}{4} \big(\bra{\phi} + \bra{\psi}\big)
        \big(\ket{\phi} + \ket{\psi}\big) \\
        & = \frac{1}{2} + \frac{1}{2} \Re \braket*{\psi}{\phi},
    \end{split}
\end{equation}
where $\Re(\cdot)$ denotes the real part of the complex number. Hence, if
we execute the quantum circuit as in Figure~\ref{fig:inner_prod} for $M$
shots and count the number of zeros, the $\Re \braket*{\psi}{\phi}$ could
be well-approximated given $M$ (which will scale as $\mathcal{O}(\frac{1}{\epsilon^2})$ for some target accuracy $\epsilon$ due to statistical sampling noise) is reasonably large.

\begin{figure}[htp]
    \centering
    \includegraphics[height=3cm]{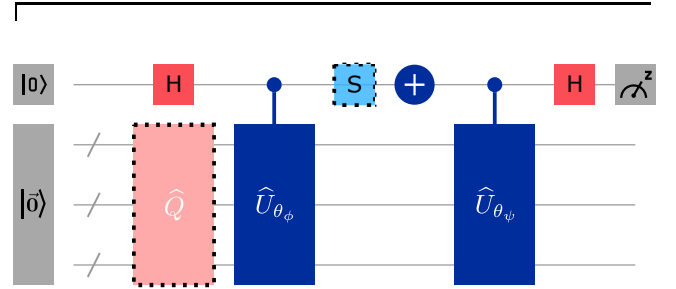}
    \caption{Extended inner product circuit.}
    \label{fig:inner_prod2}
\end{figure}

The imaginary part of $\braket*{\psi}{\phi}$ could be estimated in a
similar way. Notice that the real part in \eqref{eq:measure-real} comes
from $\braket*{\psi}{\phi} + \braket*{\phi}{\psi}$. If either $\ket{\psi}$
or $\ket{\phi}$ is replaced by $\imath \ket{\psi}$ or $\imath \ket{\phi}$,
then the real part in \eqref{eq:measure-real} corresponds to the imaginary
part of the inner product instead. One way to implement the replacement is
to add a phase gate on the ancilla qubit after stage \eqref{eq:S2}. The
resulting quantum circuit is as that in Figure~\ref{fig:inner_prod2} where
the light blue dash box is enabled. We then need to execute the circuit
for another $M$ shots to obtain the approximation of $\Im
\braket*{\psi}{\phi}$.

In addition to variational ansatz $\ket{\psi} = \Uop_{\theta_\psi}
\ket{0}$ and $\ket{\phi} = \Uop_{\theta_\phi} \ket{0}$, we could extend
the quantum circuit to variational ansatz $\ket{\psi} = \Uop_{\theta_\psi}
\Qop \ket{0}$ and $\ket{\phi} = \Uop_{\theta_\phi} \Qop \ket{0}$, where
$\Qop$ is a common quantum circuit preparing the initial state. If such
ansatzes are adopted, we could enable the light pink dash box in
Figure~\ref{fig:inner_prod2} such that the circuit $\Qop$ is applied on
$\ket{0}$ as a common circuit. The real and imaginary parts of the inner
product are evaluated similarly as before.

\subsection{$\mel*{\psi}{\Hop}{\phi}$ evaluation}

The Hamiltonian operator considered in this paper is under FCI framework
and admits the form,
\begin{equation}
    \Hop = \sum_{p,q=1}^{\Norb} h_{pq} \aop^\dagger_p \aop_q
    + \sum_{p,q,r,s=1}^{\Norb} v_{pqrs} \aop^\dagger_p \aop^\dagger_q
    \aop_s \aop_r,
\end{equation}
where $h_{pq}, v_{pqrs}$ are one-body and two-body integrals respectively.
The Hamiltonian operator $\Hop$, in general, is not a unitary operator.
Hence, it cannot be represented by a quantum circuit directly. On the
other hand, the excitation operators, $\aop^\dagger_p \aop_q$ and
$\aop^\dagger_p \aop^\dagger_q \aop_s \aop_r$, are unitary and can be
represented by quantum circuits. The detailed circuits of these excitation
operators depend on the encoding scheme. All methods proposed in this
paper do not rely on the encoding scheme. The value $\mel*{\psi}{\Hop}{\phi}$
is then calculated as a weighted summation of circuit results,
\begin{equation}
    \begin{split}
        \mel**{\psi}{\Hop}{\phi} = & \sum_{p,q=1}^{\Norb} h_{pq}
        \mel**{\psi}{\aop^\dagger_p \aop_q}{\phi} \\
        & + \sum_{p,q,r,s=1}^{\Norb} v_{pqrs}
        \mel**{\psi}{\aop^\dagger_p \aop^\dagger_q
        \aop_s \aop_r}{\phi}.
    \end{split}
\end{equation}
Thus it suffices to consider circuits for the evaluation of
$\mel*{\psi}{\Uop}{\phi}$, where $\Uop$ denotes a general unitary
operator.

\begin{figure*}[htp]
    \centering
    \includegraphics[height=3cm]{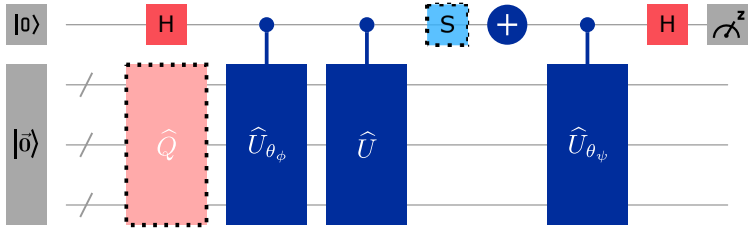}
    \caption{Extended inner product circuit with a unitary gate.}
    \label{fig:inner_prod3}
\end{figure*}

Figure~\ref{fig:inner_prod3} illustrates the circuit for
$\mel*{\psi}{\Uop}{\phi}$. This circuit can be understood in the same way
as the circuit of $\braket*{\psi}{\phi}$. The value
$\mel*{\psi}{\Uop}{\phi}$ can be explicitly written as,
\begin{equation}
    \mel**{\psi}{\Uop}{\phi} = \Big( \bra{0} \Uop_{\theta_\psi}^\dagger
    \Big) \Big( \Uop \Uop_{\theta_\phi} \ket{0}\Big) =
    \braket*{\psi}{\widetilde{\phi}},
\end{equation}
where the construction circuit for $\ket{\widetilde{\phi}}$ is $\Uop
\Uop_{\theta_\phi}$. Hence Figure~\ref{fig:inner_prod3} is essentially the
circuit in Figure~\ref{fig:inner_prod} and Figure~\ref{fig:inner_prod2}
with a modified constructing circuit for $\ket{\phi}$.

\section{Quantum Orbital Minimization Method}
\label{sec:method}

Orbital minimization method originates in the field of density functional
theory~(DFT). The minimization problem admits the following linear algebra
form,
\begin{equation} \label{eq:omm-la}
    \min_{X\in \bbC^{N \times K}} \tr \big( (2I - X^\star X) X^\star
    \Hop X \big),
\end{equation}
where $X$ is a multi-column vectors with each column denoting a state
vector, $\Hop$ is the Hamiltonian matrix, $I$ is the
identity matrix of size $K \times K$, and $\tr(\cdot)$ denotes the trace
operator.  When $\Hop$ is Hermitian negative definite, all local minima of
\eqref{eq:omm-la} are of the form~\cite{Lu2017a},
\begin{equation} \label{eq:omm-local-minima}
    X = QV,
\end{equation}
where $Q \in \bbC^{N\times K}$ are $K$ eigenvectors of $\Hop$
associated with the smallest $K$ eigenvalues, and $V \in
\bbR^{K \times K}$ is an arbitary unitary matrix. Substituting
\eqref{eq:omm-local-minima} back into \eqref{eq:omm-la}, we find that
all these local minima are of the same objective value and they are
global minima with minimizers consisting of mutually orthogonal vectors. This property has been proven analytically.\cite{Lu2017a} Unlike many other optimization problems for solving
the low-lying eigenvalue problem, OMM as in \eqref{eq:omm-la} does
not have any explicit orthogonal constraint and the orthogonality
in \eqref{eq:omm-local-minima} could be achieved without any
orthogonalization step. This property is the key reason for us to
investigate the application of \eqref{eq:omm-la} in quantum computing,
where explicit orthogonalization is difficult to implement in the
presence of hardware and environmental noise. Furthermore, despite the fact that minimizing \eqref{eq:omm-la} is a non-convex problem, its property that all local minimizers are also global minimizers which encode mutually orthogonal vectors in the case where we are minimizing over all possible vectors\cite{Lu2017a} (absent the constraint of vectors expressable as quantum circuits) makes it a compelling objective function to investigate in the quantum setting. The numerical simulations in this work serve to provide some insight into the extent to which these favorable properties in the classical setting carry over to the quantum setting where we are constrained by vectors expressable as quantum circuit ansatzes.

In the quantum computing setting, each column of $X$ is represented by the
ansatz $\ket{\psi_i} = \Uop_{\theta_i}\ket{\phi_i}$. By the nature of
quantum computing, if quantum error is not taken into account,
$\ket{\psi_i}$ is always of unit length for all $i$. Hence the $2I-X^\star
X$ term in \eqref{eq:omm-la} has one on its diagonal. Substituting the
variational ansatz into \eqref{eq:omm-la}, we obtain the optimization
problem for our quantum excited state problem,
\begin{equation} \label{eq:omm-qc}
    \min_{\theta_1, \dots, \theta_K} g(\theta_1, \dots, \theta_K)
\end{equation}
for
\begin{equation} \label{eq:omm-g}
    \begin{split}
    & g(\theta_1, \dots, \theta_K) \\
    = & \sum_{i=1}^K \mel**{\phi_i}{\Uop_{\theta_i}^\dagger \Hop
    \Uop_{\theta_i}}{\phi_i} \\
    & - \sum_{\substack{i,j=1\\i \neq j}}^K \mel**{\phi_i}{
    \Uop_{\theta_i}^\dagger \Uop_{\theta_j}}{\phi_j}
    \mel**{\phi_j}{\Uop_{\theta_j}^\dagger \Hop
    \Uop_{\theta_i}}{\phi_i}. \\
    \end{split}
\end{equation}

If the variational space is rich enough such that vectors in $QV$ as in
\eqref{eq:omm-local-minima} can be well-approximated by $\big\{
\Uop_{\theta_i}\ket{\phi_i} \big\}_{i=1}^K$, then qOMM can achieve a good
approximation to the global minimum of its linear algebra counterpart
\eqref{eq:omm-la}. In practice, if we obtain the global minimum value of
\eqref{eq:omm-la} for qOMM (though the value is not known in advance),
then we are confident that the ansatz is rich enough and the optimizer
works perfectly in finding the global minimum for parameters $\theta_1,
\dots, \theta_K$. However, we lack the theoretical understanding of the
energy landscape of \eqref{eq:omm-qc}. In part, the dependence of $g$ on
$\theta_1, \dots, \theta_K$, like that in the deep neural network, is
highly nonlinear and nonconvex. It could be analyzed for some simple
ansatz circuit. However, for most complicated circuits, the analysis
remains difficult. On the other hand, from the linear algebra point of
view, when the variational space is rich enough, we could view the problem
as optimizing the vectors $X$ directly in \eqref{eq:omm-la}. The nature of
quantum computer poses unit length constraints on each column of $X$.
Adding extra unit length constraints would dramatically change the energy
landscape of the original problem \eqref{eq:omm-la}. We defer the
theoretical understanding of \eqref{eq:omm-qc} to future work. In this
paper, we focus on the numerical performance of \eqref{eq:omm-qc}.

For numerical optimization of the objective function $g$, note that its
gradient with respect to a parameter $\theta_{i \alpha}$ is given by 
\begin{widetext}
\begin{equation} \label{eq:grad}
    \begin{split}
        \frac{\diff g}{\diff \theta_{i\alpha}} = &
        2 \Re \mel**{\phi_i}{\frac{\diff \Uop^\dagger_{\theta_i}}{\diff
        \theta_{i\alpha}} \Hop \Uop_{\theta_j}}{\phi_j}
        - 2 \Re \sum_{\substack{j=1\\j \neq i}}^K
        \mel**{\phi_i}{\frac{\diff \Uop^\dagger_{\theta_i}}{\diff
        \theta_{i\alpha}} \Uop_{\theta_j}}{\phi_j}
        \mel**{\phi_j}{\Uop_{\theta_j}^\dagger \Hop \Uop_{\theta_i}}{\phi_i}\\
        & - 2 \Re \sum_{\substack{j=1\\j \neq i}}^K
        \mel**{\phi_i}{\Uop^\dagger_{\theta_i}
        \Uop_{\theta_j}}{\phi_j}
        \mel**{\phi_j}{\Uop_{\theta_j}^\dagger \Hop
        \frac{\diff \Uop_{\theta_i}}{\diff \theta_{i\alpha}}}{\phi_i},\\
    \end{split}
\end{equation}
\end{widetext}

\noindent
where $\theta_{i\alpha}$ is the $\alpha$-th parameter in $\theta_i$.
From \eqref{eq:grad}, we notice that the gradient evaluation requires an
efficient quantum circuit for $\frac{\diff \Uop_{\theta_i}}{\diff
\theta_{i\alpha}}$. The existence of such quantum circuits is ansatz
dependent. Assume that we can evaluate \eqref{eq:grad} for all
parameters.Then the update on parameters follows a gradient descent method
as
\begin{equation*}
    \theta^{(t+1)} = \theta^{(t)} - \tau \nabla_\theta g(\theta^{(t)}),
\end{equation*}
for $\theta^{(t+1)}$ and $\theta^{(t)}$ being all parameters at
$(t+1)$-th and $(t)$-th iteration respectively, and $\tau$ is the
stepsize. Since all evaluations of quantum circuits involve randomness,
in practice, this amounts to using the stochastic gradient descent
method. Other advanced first order methods recently developed in deep
neural network literature, \eg, ADAM, AdaGrad, coordinate descent
method, etc., could be used to accelerate. These methods could achieve
better performance than the vanilla stochastic gradient descent method.

When the gradient circuit is not available, we can adopt zeroth-order
optimizers to solve \eqref{eq:omm-qc}. In this work, we use L-BFGS-B\cite{doi:10.1137/0916069} as the optimizer,
which uses a 2-point finite difference gradient approximation and thus does not require explicit implementation of the circuit for $\frac{\diff \Uop_{\theta_i}}{\diff
\theta_{i\alpha}}$. We find
this optimizer to work well in the noiseless simulations we consider here,
but we note that other promising candidates for noisy simulations and
experimental calculations exist, which could also be used for our proposed
objective function. In particular, using simultaneous perturbation
stochastic approximation~(SPSA) for small molecule VQE calculations has
been demonstrated experimentally on superconducting
hardware~\cite{Kandala2017} and a quantum natural gradient version of SPSA
also exists~\cite{Gacon2021}.

When the optimization problem is solved, we will obtain a set of
parameters $\theta_1, \dots, \theta_K$ such that
$\Uop_{\theta_1}\ket{\phi_1}, \dots, \Uop_{\theta_K}\ket{\phi_K}$
approximate vectors in $QV$ as in \eqref{eq:omm-local-minima}.
Numerically, we find that $V$ is very close to an identity matrix, while a
post-processing procedure still improves the accuracy of the
approximation. The post-processing procedure is slightly different from
that in MCVQE. We construct two matrices with their $(i,j)$-th elements
being,
\begin{equation}
    \begin{split}
        A_{ij} = \mel**{\phi_i}{\Uop_{\theta_i}^\dagger \Hop
        \Uop_{\theta_j}}{\phi_j} \text{ and}\\
        B_{ij} = \mel**{\phi_i}{\Uop_{\theta_i}^\dagger
        \Uop_{\theta_j}}{\phi_j}.\\
    \end{split}
\end{equation}
Then a generalized eigenvalue problem with matrix pencil $(A,B)$ is
solved as,
\begin{equation}
    AR = BR\Lambda,
\end{equation}
where $R$ denotes the eigenvectors and $\Lambda$ is a diagonal matrix with
eigenvalues of $(A,B)$ being its diagonal entries. The desired eigenvalues
of $\Hop$ are in $\Lambda$ and the corresponding eigenvectors admits
$\sum_{j=1}^K\Uop_{\theta_j} \ket{\phi_j} R_{ji}$ for $i=1,\dots, K$. The
eigenvector, which is the excited state of $\Hop$, can be explicitly
constructed as a linear combination of unitary operators~\cite{Childs2012}
on a quantum computer.

qOMM has several unique features compared to existing excited state vector methods. The targets of folded
spectrum method~\cite{McClean2016} and witnessing eigenspectra
solver~\cite{Santagati2017,Santagati2018} are different from that of
qOMM. Quantum deflation method~\cite{Higgott2019, Jones2019} addresses
the excited states one-by-one, and the optimization problems therein are
of different difficulty as that in qOMM. The most related excited state
vector methods to qOMM are SSVQE and MCVQE. Since SSVQE and MCVQE are very
similar to each other, we focus on the difference among qOMM and SSVQE in this
paper. Both the qOMM and weighted SSVQE algorithms~\cite{Nakanishi2019} take
a set of input states and seek to simultaneously optimize the parameterized
circuit and obtain the low-lying eigenvalues and eigenvectors of the Hermitian
operator. Both methods achieve their goal via solving a single optimization
problem. There are two major differences between them. The first one is how
the mutual orthogonality of the eigenvectors is enforced. The construction of
the weighted SSVQE objective function ensures that the set of states being
evaluated is mutually orthogonal throughout optimization iteration. The
proposed qOMM implicitly embeds this constraint into the objective function
such that the states being evaluated are only guaranteed to be mutually
orthogonal at the global minimum. The first difference leads to the second
major difference. SSVQE adopts the ansatz circuits for different states that
must admit mutual orthogonal property. Hence, they restrict their ansatz
circuits to be a set of identical parameterized circuits acting on mutually
orthogonal initial states. qOMM does not constrain the ansatz circuits in this
way. qOMM could adopt as its set of parameterized circuits any combination of
ansatz circuits that have been tested in the literature, \eg, ansatz circuits
in MCVQE~\cite{Parrish2019} and SSVQE~\cite{Nakanishi2019}, ansatz circuits
in deflation method~\cite{Higgott2019, Jones2019}, etc. In addition to these
ansatz circuits, qOMM could also adopt other ansatz circuits not been tested
by excited state methods. In this work, we adopt UCCSD\cite{Romero_2018}
as our main ansatz circuit so that the particle number is preserved throughout.

\section{Numerical Results}
\label{sec:num}

We present the numerical results of our proposed method obtained from a
classically simulated quantum computer. We apply it to the problem
of finding low-lying eigenvalues of the electronic structure Hamiltonian
for near-equilibrium configurations of two different molecules as
well the toy model consisting of 4 hydrogen atoms arranged in a square lattice. The numerical tests in this section serve to explore
what advantages or disadvantages may arise for each of the qOMM and weighted
SSVQE approaches (using weight vectors of the form [$n$, $n-1$, ..., $1$] for
finding $n$ states) in different situations. Section~\ref{sec:H2andH4} presents the results for \ch{H_{2}} and the hydrogen square model obtained from noiseless simulations. Section~\ref{sec: NoisyH2} presents the results obtained from simulating \ch{H_{2}} with a depolarizing noise model. Section~\ref{sec: LiH} presents the results for \ch{LiH} obtained from noiseless simulations.

All codes for these simulations are implemented using the Qiskit~\cite{Qiskit}
library. The electronic structure Hamiltonians were generated using molecular
data from PySCF~\cite{Sun2018} in the STO-3G basis and mapped to qubit
Hamiltonians using the Jordan-Wigner mapping for the noiseless simulations and the Parity mapping for the noisy simulations. The relative accuracies of the
results for both methods are calculated by comparing them to their numerically
exact counterparts obtained by numerically exact diagonalization of the
Hamiltonians. Circuits involved in noiseless qOMM simulations were
conducted using Qiskit's \emph{StateVector} simulator. Circuits involved in
noiseless SSVQE simulations were conducted using Qiskit's \emph{QasmSimulator}
in conjunction with the \emph{AerPauliExpectation} method for computing
expectation values in order to improve runtime performance. Both of these
methods produce ideal, noiseless results. Simulations for finding the
energy levels of \ch{H2} were performed at an interatomic distance of
$0.735$ Angstroms, those of \ch{LiH} were performed at an interatomic
distance of $1.595$ Angstroms, and those of the hydrogen square lattice model were performed at an interatomic distance of $1.23$ Angstroms. Simulation results for the hydrogen square model and \ch{H_{2}} at stretched bond distances can also be found in Appendix~\ref{appendix: Convergence Plots}. For the \ch{LiH} simulations,
we freeze the two core electrons to reduce the problem size from 12 qubits
to 10 qubits.

\begin{table*}[htb]
    \centering
    \begin{tabular}{ccccccccc}
        \toprule
        Molecule & Algorithm & \multicolumn{7}{c}{UCCSD Repetitions} \\
        \cmidrule{3-9}
        & & 1-rep & 2-rep & 3-rep & 4-rep & 5-rep & 6-rep & 7-rep \\
        \toprule
        \multirow{2}{*}{\ch{H2} (0.735 {\AA})}
        & qOMM (3 states) & 100\% & $-$ & $-$ & $-$ & $-$ & $-$ & $-$ \\
        & SSVQE (3 states) & 0\% & 70\% & 100\% & $-$ & $-$ & $-$ & $-$ \\
        \toprule
        \multirow{2}{*}{\ch{H2} (1.47 {\AA})}
        & qOMM (3 states) & 100\% & $-$ & $-$ & $-$ & $-$ & $-$ & $-$ \\
        & SSVQE (3 states) & 0\% & 90\% & 100\% & $-$ & $-$ & $-$ & $-$ \\
        \toprule
        \multirow{4}{*}{Hydrogen square (1.23 {\AA})}
        & qOMM (2 states)  & 0\% & 100\% & $-$ & $-$ & $-$ & $-$ & $-$ \\
        & qOMM (3 states)  & 0\% & 100\% & $-$ & $-$ & $-$ & $-$ & $-$ \\
        & SSVQE (2 states) & $-$ & 0\% & 100\% & $-$ & $-$ & $-$ & $-$ \\
        & SSVQE (3 states) & $-$ & $-$ & 0\% & 100\% & $-$ & $-$ & $-$ \\
        \toprule
        \multirow{4}{*}{Hydrogen square (2.46 {\AA})}
        & qOMM (2 states)  & 0\% & 90\% & $-$ & $-$ & $-$ & $-$ & $-$ \\
        & qOMM (3 states)  & 0\% & 100\% & $-$ & $-$ & $-$ & $-$ & $-$ \\
        & SSVQE (2 states) & $-$ & 0\% & 100\% & $-$ & $-$ & $-$ & $-$ \\
        & SSVQE (3 states) & $-$ & $-$ & 0\% & 100\% & $-$ & $-$ & $-$ \\
        \toprule
        \multirow{12}{*}{\ch{LiH} (1.595 {\AA})}
        & qOMM (2 states)  & 0\% & 100\% & $-$ & $-$ & $-$ & $-$ & $-$ \\
        & qOMM (3 states)  & 0\% & 100\% & $-$ & $-$ & $-$ & $-$ & $-$ \\
        & qOMM (4 states)  & $-$ & 100\% & $-$ & $-$ & $-$ & $-$ & $-$ \\
        & qOMM (5 states)  & $-$ & 100\% & $-$ & $-$ & $-$ & $-$ & $-$ \\
        & qOMM (6 states)  & $-$ & 100\% & $-$ & $-$ & $-$ & $-$ & $-$ \\
        & qOMM (7 states)  & $-$ & 100\% & $-$ & $-$ & $-$ & $-$ & $-$ \\
        & SSVQE (2 states) & 0\% & 80\%  & $-$ & $-$ & $-$ & $-$ & $-$ \\
        & SSVQE (3 states) & 0\% & 0\%   & 70\% & $-$ & $-$ & $-$ & $-$ \\
        & SSVQE (4 states) & $-$ & $-$ & 0\% & 100\% & $-$ & $-$ & $-$ \\
        & SSVQE (5 states) & $-$ & $-$ & $-$ & 10\% & 100\% & $-$ & $-$ \\
        & SSVQE (6 states) & $-$ & $-$ & $-$ & $-$ & 10\% & 100\% & $-$ \\
        & SSVQE (7 states) & $-$ & $-$ & $-$ & $-$ & $-$ & 60\% & 100\% \\
        \bottomrule
    \end{tabular}
    \caption{The success rate of given problem instances converging to
    within a relative accuracy of $10^{-5}$ for a given number of UCCSD
    repetitions using a randomized ansatz parameter initialization.}
    \label{tab:success_rate_table}
\end{table*}

The set of initial states for both algorithms and all molecular
Hamiltonians were chosen to be the Hartree-Fock state and low-lying
single-particle excitations above it. For example, if we wish to
find three energy levels for 4-qubit \ch{H2}, in the Jordan-Wigner mapping this would be the set:
\{$\ket{01}_\alpha\ket{01}_\beta$, $\ket{01}_\alpha\ket{10}_\beta$,
$\ket{10}_\alpha\ket{01}_\beta$\}. For LiH, this would be the set:
\{$\ket{00001}_\alpha\ket{00001}_\beta$, $\ket{00001}_\alpha\ket{00010}_\beta$,
$\ket{00010}_\alpha\ket{00001}_\beta$\}, where the subscript $\alpha$
and $\beta$ denote the spin group~\footnote{In the Qiskit implementation
of $n$-qubit basis states in the Jordan-Wigner representation, the first
$\frac{n}{2}$ qubits encode spin-up ($\alpha$) and the second $\frac{n}{2}$
qubits encode spin-down ($\beta$). Thus we have chosen these three basis
states because they are elements of the 2-particle, spin magnetization $S_z$
= 0 subspace of the full $2^{n}$-dimensional Fock space.}. Given that the
UCCSD\cite{Romero_2018} ansatz preserves the numbers of spin-up and spin-down,
the above choice of initial states constrains these variational quantum
algorithms to search only the desired subspace, lessening the computational
difficulty and producing physically meaningful results at the same time. For
this reason, UCCSD is our ansatz of choice for these simulations. If we wish
to find only two energy levels, we would omit the highest-energy state from
the set. This choice of the set of states allows us to study two different
types of algorithm initializations: one in which all of the ansatz parameters
are randomly sampled according to a uniform distribution on $[-2\pi, 2\pi)$
and another one in which they are all set to zero. With the random parameter
initialization, we can study the robustness of the algorithms with respect
to their starting point in the parameter space.  With the ``zero vector''
initialization, the UCCSD ansatz circuit is initialized to the identity
and the states after the initial ansatz circuit application remain the
Hartree-Fock state and low-lying single-particle excitations above the
Hartree-Fock states. For most molecules, these states from Hartree-Fock
calculation are good approximations of the ground state and low-lying excited
states under FCI.  The optimization from the ``zero vector'' initialization
could be viewed as a local optimization and improves the performance of
optimizers upon the random initialization. In Qiskit's implementation of
ansatz circuits such as UCCSD, one can increase the expressiveness of the
ansatz by repeating the corresponding quantum circuit block pattern $r$
times, which comes at the cost of both the circuit depth and the number
of variational parameters increasing by a factor of $r$. In this paper, we
refer to an ansatz circuit that consists of the UCCSD circuit block pattern
repeated $r$ times as $r$-repetition UCCSD (or simply $r$-UCCSD). We run both
algorithms for several different numbers of repetitions in order to account
for the fact that the necessary ansatz expressiveness for convergence is not
known \emph{a priori}. Such a study allows us to explore the dependence of
convergence success on the circuit depth for each algorithm.

When using a random initialization, we run each algorithm ten times for each
setting. In Sections~\ref{sec:H2andH4} and~\ref{sec: LiH}, for each setting we plot one run roughly representative of the average convergence. In the noisy results in Section~\ref{sec: NoisyH2}, we plot the run which obtained the lowest objective function convergence. The complete set of all ten runs for all of these simulations are plotted in the appendix and their convergence success rates are summarized in Table~\ref{tab:success_rate_table} for completeness. When using the
``zero vector'' initialization, we run each algorithm only once. Running
the algorithm multiple times for this initialization is neither necessary
nor useful in the absence of noise because the outcome is deterministic for
a given initial point.

\subsection{\ch{H2} and Hydrogen square}\label{sec:H2andH4}

%Table was originally put here- for future reference.

\begin{figure*}[h!]
    \centering
    \begin{subfigure}{0.49\linewidth}
        \centering
        \includegraphics[width=\linewidth]{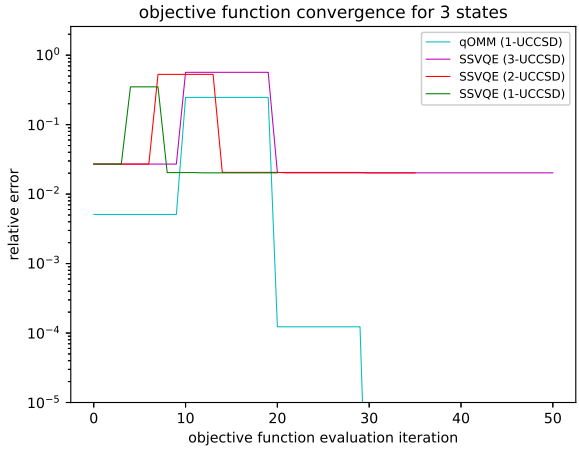}
        \caption{Zero vector initialization}
        \label{fig:H2-3states-one-curve-zero-vector}
    \end{subfigure}
    \begin{subfigure}{0.49\linewidth}
        \centering
        \includegraphics[width=\linewidth]{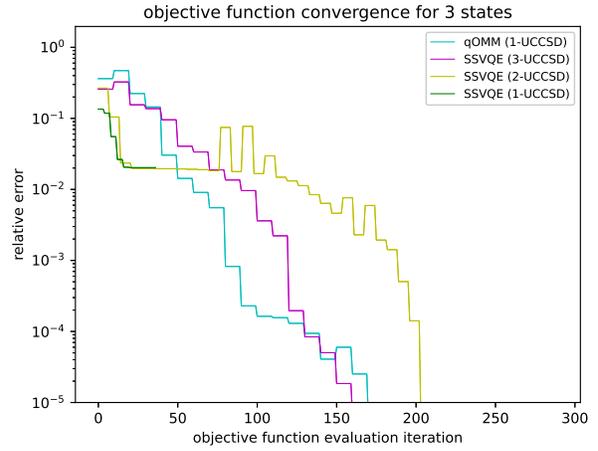}
        \caption{Random initialization}
        \label{fig:H2-3states-one-curve-random}
    \end{subfigure}
    \caption{Convergence (noise-free) of  the relative error $\frac{\abs{f_{i} - f_{exact}}}{\abs{f_{exact}}}$ of qOMM and SSVQE for 3 \ch{H_2} states at an interatomic distance of 0.735 {\AA}.}
    \label{fig:H2-one-curve}
\end{figure*}

We begin by presenting the results for the 4 and 8 qubit systems we consider here:
\ch{H2} and the hydrogen square model. Figure~\ref{fig:H2-one-curve} illustrates the convergence of the objective
function for each algorithm when attempting to calculate the first three
energy levels of the \ch{H2} molecule at the equilibrium bond distance.

\begin{figure*}[h!]
    \centering
    \begin{subfigure}{0.49\linewidth}
        \centering
        \includegraphics[width=\linewidth]{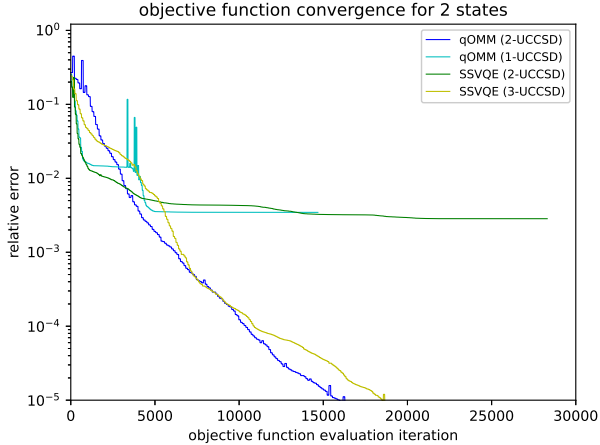}
        \caption{Two states}
        \label{fig:H4-2-states-one-curve}
    \end{subfigure}
    \begin{subfigure}{0.49\linewidth}
        \centering
        \includegraphics[width=\linewidth]{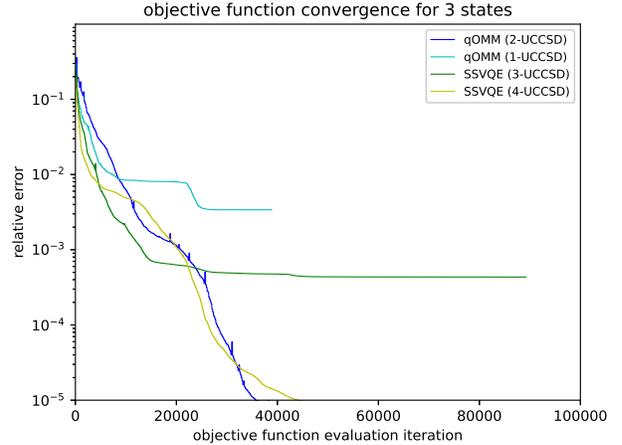}
        \caption{Three states}
        \label{fig:H4-3-states-one-curve}
    \end{subfigure}
    \caption{Convergence (noise-free) of  the relative error $\frac{\abs{f_{i} - f_{exact}}}{\abs{f_{exact}}}$ of qOMM and SSVQE for the hydrogen square model at an interatomic distance of 1.23 {\AA} using a random parameter initialization.}
    \label{fig:H4-one-curve}
\end{figure*}

Figure~\ref{fig:H2-3states-one-curve-random} demonstrates how qOMM and SSVQE both
converge to the global minimum, but notably, SSVQE cannot do so with just
1-repetition UCCSD, requiring two repetitions in order to converge to
within a relative accuracy of $10^{-5}$. The success rate is further
reported in Table~\ref{tab:success_rate_table}.
Figure~\ref{fig:H2-3states-one-curve-zero-vector} illustrates an attempt to solve the same
problem, except the ansatz parameters are all initialized to zero. In this
instance, we see that qOMM converges more quickly by a factor of roughly 5 compared to its random initialization counterpart, while SSVQE
gets stuck in a local minimum and does not converge regardless of the
ansatz circuit depth used. In both figures and all later convergence
figures in this paper, we depict the convergence curve as the relative
error (defined as $\frac{\abs{f_{exact} - f_{measured}}}{\abs{f_{exact}}}$) against the number of the objective function evaluations, which is
considered the most expensive operation in VQE-type algorithms. The L-BFGS-B implementation used in this paper uses a 2-point finite difference method such that each parameter is perturbed slightly from a given reference point common to all of them. Thus for an $n$-parameter objective function, $n+1$ function evaluations are required for each parameter update step in order to estimate the gradient. Hence, in all convergence figures, we observe stair-like curves of width $n+1$.

Figure~\ref{fig:H4-one-curve} illustrates the convergence results for the hydrogen square lattice at an interatomic distance of $1.23$ Angstroms. We can see from these figures that similarly to \ch{H_{2}}, the difference between SSVQE and qOMM is primarily in the number of UCCSD repetitions needed to converge. When calculating 2 states, qOMM requires 2-UCCSD whereas SSVQE requires 3-UCCSD. When calculating 3 states, qOMM still requires only 2-UCCSD, whereas SSVQE requires 4-UCCSD. These observations are further tabulated in Table \ref{tab:success_rate_table}.

\subsection{Noisy \ch{H_{2}} simulations}\label{sec: NoisyH2}

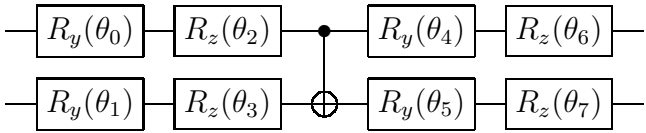
\begin{figure}[H]
    \centering
    \centerline{\Qcircuit @C=1em @R=.7em {
                        & \gate{R_{y}(\theta_{0})} & \gate{R_{z}(\theta_{2})} & \ctrl{1} & \gate{R_{y}(\theta_{4})} & \gate{R_{z}(\theta_{6})} & \qw \\
                        & \gate{R_{y}(\theta_{1})} & \gate{R_{z}(\theta_{3})} & \targ & \gate{R_{y}(\theta_{5})} & \gate{R_{z}(\theta_{7})}  & \qw
}}
    \caption{Base circuit block pattern used for noisy \ch{H_{2}} simulations.}
    \label{fig:NoisyH2Ansatz}
\end{figure}

We now run \ch{H_{2}} simulations in the presence of a classically
simulated noise model on Qiskit's \emph{AerSimulator}. Each one-qubit gate
in the circuit is modelled as being accompanied by a local depolarizing
channel with some probability of error $p_{error}$. Two-qubit gates are
accompanied by a tensor product of two local depolarizing channels acting on
each qubit. No error mitigation strategies are employed. We use the Parity mapping in order to use symmetry considerations to reduce
the \ch{H_{2}} Hamiltonian to a 2 qubit representation\cite{https://doi.org/10.48550/arxiv.1701.08213}. We construct
ansatzes for each algorithm using the circuit block pattern shown in Figure~\ref{fig:NoisyH2Ansatz}.

\begin{figure*}[h!]
    \centering
    \begin{subfigure}{0.49\linewidth}
        \centering
        \includegraphics[width=\linewidth]{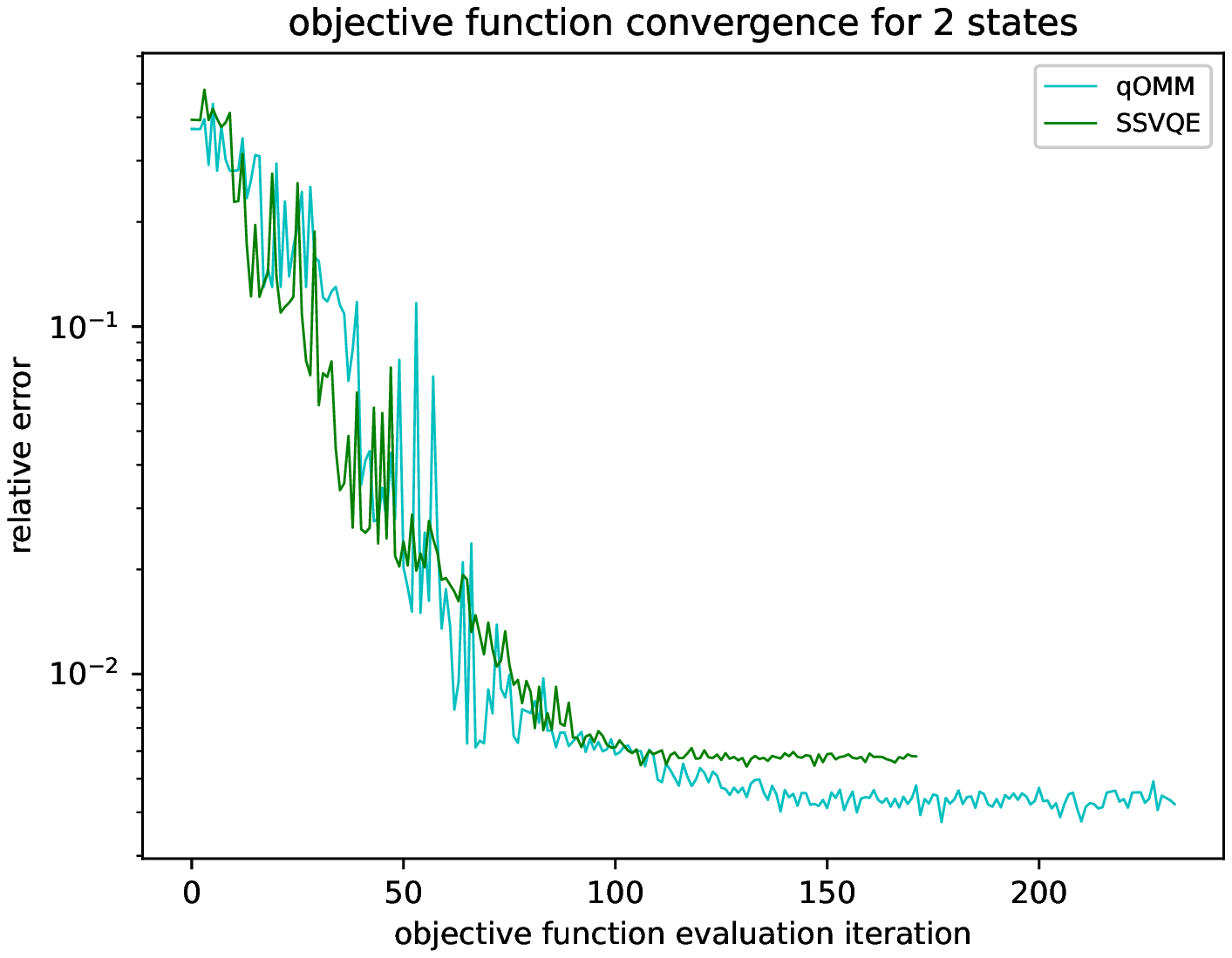}
        \caption{Two states}
        \label{fig:H2-2states-noisy-one-curve}
    \end{subfigure}
    \begin{subfigure}{0.49\linewidth}
        \centering
        \includegraphics[width=\linewidth]{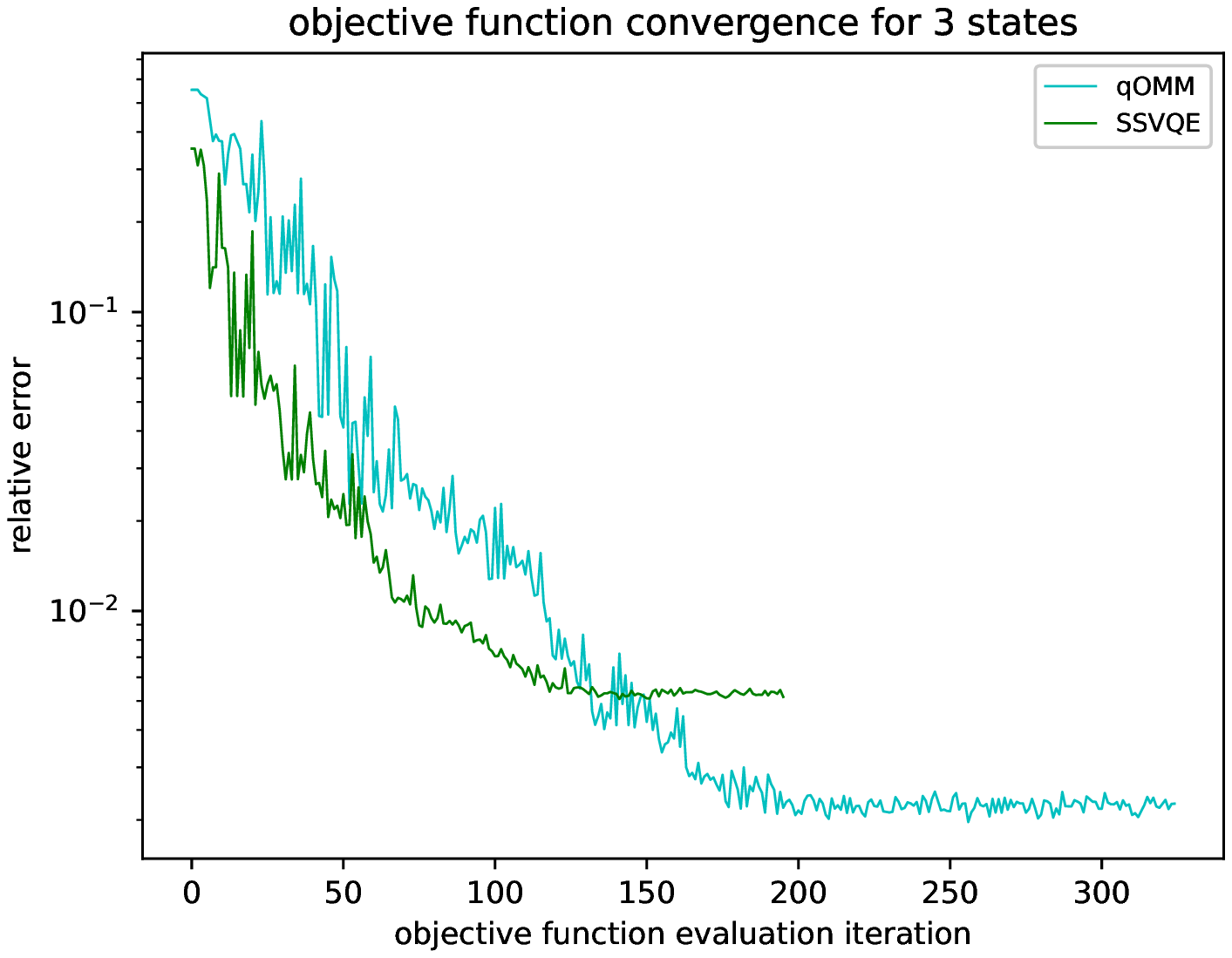}
        \caption{Three states}
        \label{fig:H2-3states-noisy-one-curve}
    \end{subfigure}
    \caption{Convergence of  the relative error $\frac{\abs{f_{i} - f_{exact}}}{\abs{f_{exact}}}$ of qOMM and SSVQE for \ch{H_2} at an interatomic distance of 0.735 {\AA}, where each circuit gate is modelled as having a probability of local depolarizing error $p_{error} = 0.001$.}
    \label{fig:H2-noisy-one-curve}
\end{figure*}

This pattern can be repeated an arbitrary number of
times to construct increasingly expressive ansatzes in the same way that the
number of UCCSD repetitions was varied in the noiseless simulations. The
Qiskit compiler is used to compile all circuits to the set of basis gets
consisting of $R_{z}$, $R_{y}$, CNOT, $\sqrt{X}$, and the identity. For all
problem instances, we use the minimum number of repetitions necessary to
converge in the absence of noise in order to ensure that we are not simply
measuring the ability of the ansatz to represent the global minimum. This
corresponds to one repetition for qOMM and two repetitions for SSVQE. We use COBYLA as the classical optimization subroutine. This
optimizer is more suitable for noisy simulations than L-BFGS-B due to
its lack of a need to calculate gradient information. $10^6$ circuit
samples are used to evaluate all inner product terms and expectation
values. $p_{error}$ is set to $10^{-3}$. Both SSVQE and qOMM are run ten
times with randomly initialized parameters. The convergence of the runs which achieved the lowest objective function values are given in Figure~\ref{fig:H2-noisy-one-curve}. The results for all ten runs are
given in Appendix~\ref{appendix: Convergence Plots}. We can see from these figures that both SSVQE and qOMM demonstrate a robustness to noise, although they do not achieve the same accuracy as the noiseless results.

\subsection{\ch{LiH}}\label{sec: LiH}

\begin{figure*}[h!]
    \centering
    \begin{subfigure}{0.49\linewidth}
        \centering
        \includegraphics[width=\linewidth]{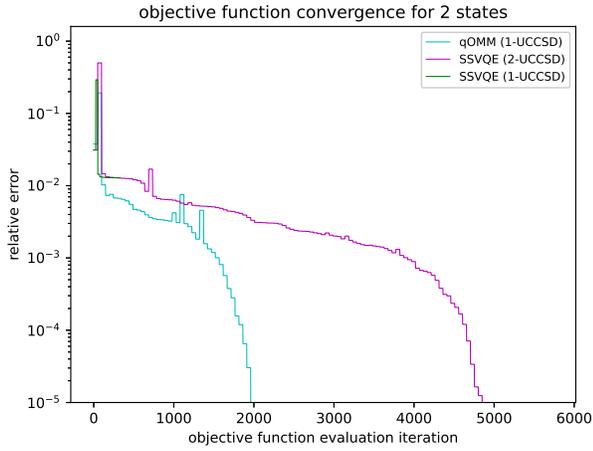}
        \caption{Two states (zero vector initialization)}
        \label{fig:LiH-2states-zero}
    \end{subfigure}
    \begin{subfigure}{0.49\linewidth}
        \centering
        \includegraphics[width=\linewidth]{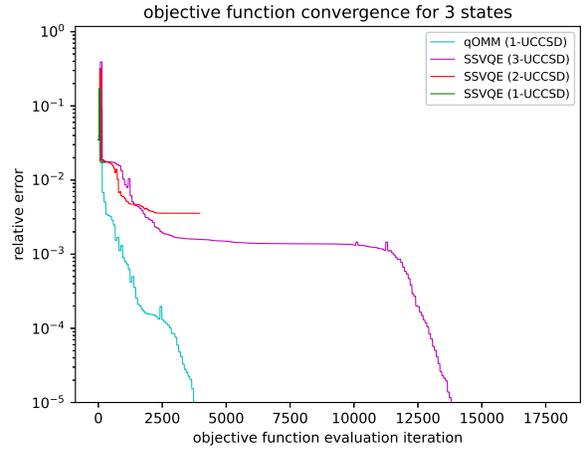}
        \caption{Three states (zero vector initialization)}
        \label{fig:LiH-3states-zero}
    \end{subfigure}
    \begin{subfigure}{0.49\linewidth}
        \centering
        \includegraphics[width=\linewidth]{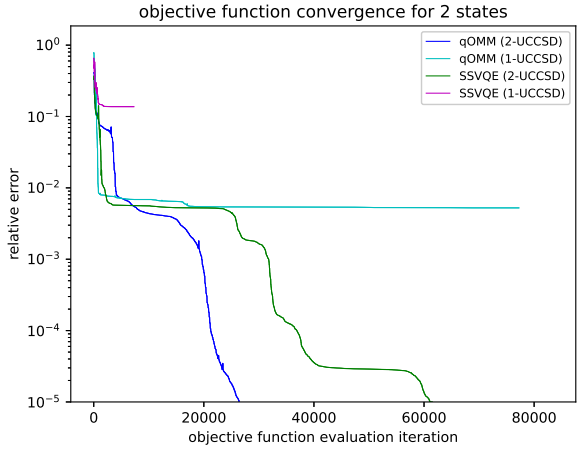}
        \caption{Two states (random initialization)}
        \label{fig:LiH-2states-one-curve-random}
    \end{subfigure}
    \begin{subfigure}{0.49\linewidth}
        \centering
        \includegraphics[width=\linewidth]{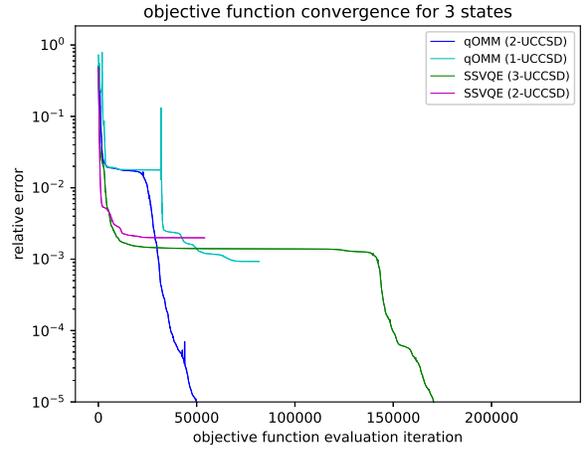}
        \caption{Three states (random initialization)}
        \label{fig:LiH-3states-one-curve-random}
    \end{subfigure}
    \caption{Convergence of the relative error $\frac{\abs{f_{i} - f_{exact}}}{\abs{f_{exact}}}$ of qOMM and SSVQE for \ch{LiH}.}
    \label{fig:LiH-one-curve-plots}
\end{figure*}

\begin{figure}[h!]
    \centering
    \includegraphics[width=\linewidth]{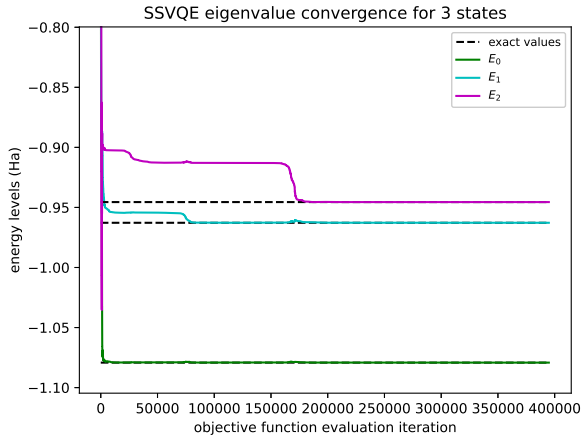}
    \caption{Convergence of the first three energy levels of \ch{LiH} for SSVQE
    using a random parameter initialization.}
    \label{fig:LiH-3states-SSVQE}
\end{figure}

\begin{figure}[htb]
    \centering
    \includegraphics[width=\linewidth]{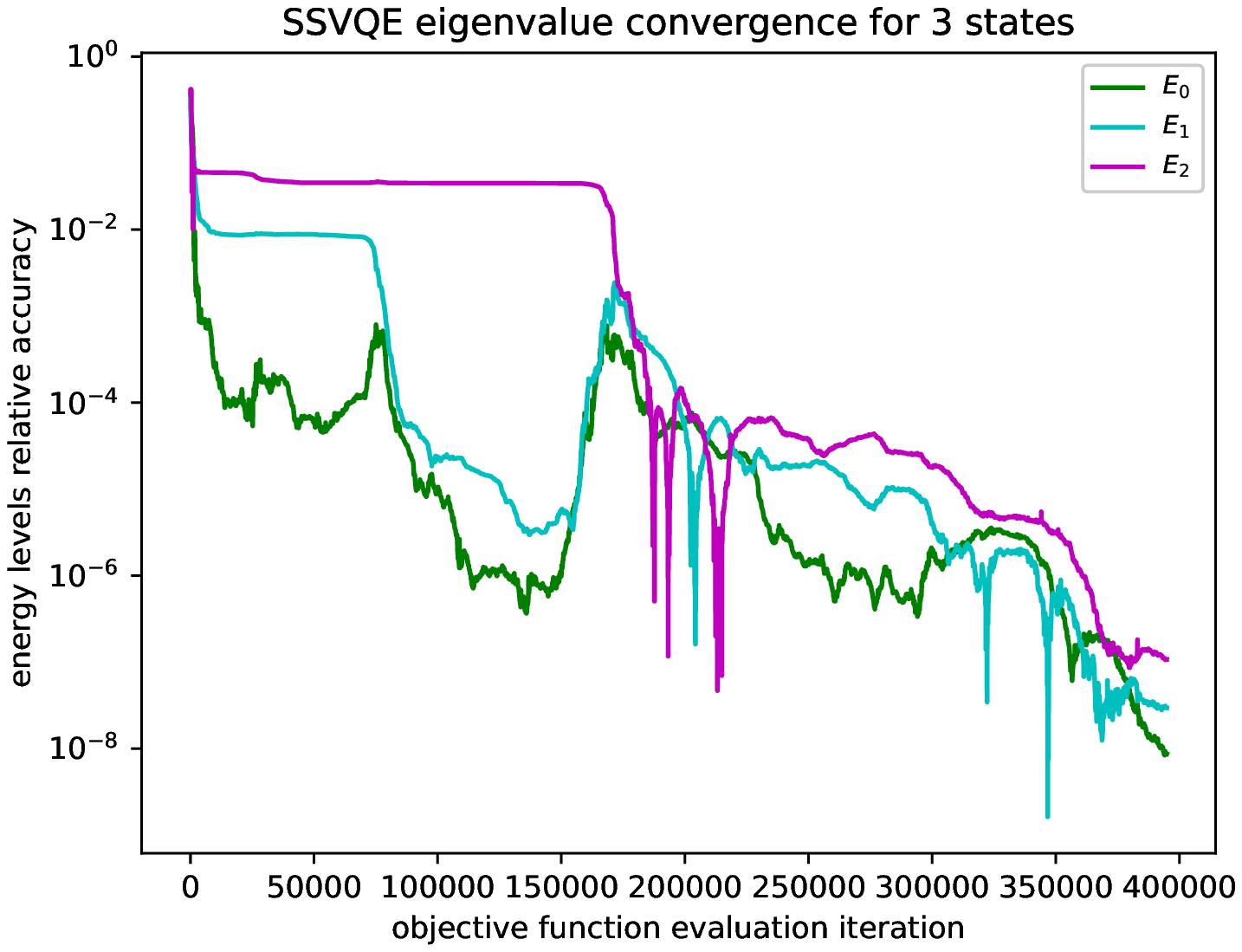}
    \caption{Relative accuracy $\frac{\abs{E_{n,i} - E_{n,exact}}}{\abs{E_{n,exact}}}$ of the convergence of the first three
    energy levels of \ch{LiH} for SSVQE using a
    random parameter initialization.}
    \label{fig:LiH-3states-accuracy-SSVQE}
\end{figure}

We now present the results for the 10-qubit system we consider here:
\ch{LiH}. Figure~\ref{fig:LiH-one-curve-plots} illustrates the
convergence of the objective function for both algorithms with various UCCSD
ansatzes when calculating up to
the first three energy levels. We can see from Figure~\ref{fig:LiH-2states-one-curve-random} that when the ansatz
parameters are randomly initialized, both algorithms demonstrate the
ability to converge within an accuracy of $10^{-5}$, but require
2-repetition UCCSD to do so. From Figure~\ref{fig:LiH-2states-zero} we can
see that when all of the parameters are initialized to zero, both
algorithms converge to within a relative accuracy of $10^{-5}$ much more
quickly than their randomly initialized counterparts. Notably, qOMM
requires only 1-repetition UCCSD to achieve this convergence, while SSVQE
requires two repetitions. When three states are calculated, qOMM can
achieve a convergence below $10^{-5}$ with 2-repetition UCCSD, whereas
SSVQE requires three repetitions. From Figure~\ref{fig:LiH-3states-zero}
we can see that when the ansatz parameters are all initialized to zero,
both algorithms quickly converge to the global minimum. Notably, qOMM
requires only 1-UCCSD repetition with this initialization. When attempting to find larger numbers of states, we can see from Table~\ref{tab:success_rate_table} that qOMM can converge with 2-UCCSD for all numbers of states considered, whereas SSVQE requires an increasing number of ansatz repetitions to find increasing numbers of states. In the following, we discuss the numerical results on \ch{LiH} mainly from three perspectives: convergence plateau, initialization, and ansatz circuit depth.

\emph{Convergence Plateau.} In many of the \ch{LiH} convergence plots we show here, both qOMM and SSVQE initially
converge at a rapid rate but then hit a plateau and stall the convergence
until they escape from the ``bad'' region of the energy landscape. As
long as the algorithm escapes from the ``bad'' region, the convergence
rate resumes being rapid. These ``bad'' regions of the energy landscape
(regardless of what their nature may be) seem to be the main obstacle for
the convergence of both algorithms, which is more severe to SSVQE. Studying
this apparent feature of the energy landscape may provide some insight as
to how both algorithms may be improved and what limitations are inherent
to the construction of the cost functions. From a numerical linear algebra
point of view, such a convergence behavior is not very surprising. The
objective function in SSVQE is convex, while with the consideration of the
orthogonality constraint and the parameterization of the ansatz circuits,
the energy landscape of SSVQE becomes non-convex. On the other hand, the
objective function of qOMM, by itself, is non-convex but has no spurious
local minima~\cite{Lu2017a, Gao2020, Gao2021}. When the parameterization
of the ansatz circuits is taken into consideration, the energy landscape
of qOMM is non-convex and could have spurious local minima. For non-convex
energy landscapes, usual optimizers, including L-BFGS-B, are efficient
in a neighborhood of local minima whereas, around strict saddle points,
without second-order information, the optimizers could stall there for a
long time. From a practical point of view, we can further investigate the
convergence of each eigenvalue. Since the issue is more severe for SSVQE, we
focus on the SSVQE rather than qOMM. The construction of the SSVQE objective
function allows us to obtain the estimated values for each eigenvalue at
every function evaluation (in addition to the objective function as a whole)
at a negligible additional computational cost. This allows us to gain some
intuition as to why the SSVQE objective function convergence is observed
to plateau in many of the \ch{LiH} runs before steeply converging below a
relative accuracy of $10^{-5}$. In Figure~\ref{fig:LiH-3states-SSVQE} and
Figure~\ref{fig:LiH-3states-accuracy-SSVQE}, we plot the eigenvalues and their
relative accuracies, respectively, for one of the ten randomly-initialized
SSVQE runs. We can see from
Figure~\ref{fig:LiH-3states-SSVQE} that while $E_{0}$ converges rapidly, the
convergence of $E_{1}$ and $E_{2}$ plateau and do not converge for many more
function evaluations. From Figure~\ref{fig:LiH-3states-accuracy-SSVQE} we can
see that $E_{1}$ does not escape from its plateau until the accuracy of $E_{0}$
is sufficiently decreased. This is followed by the accuracies of $E_{0}$
and $E_{1}$ increasing together. $E_{2}$ does not escape from its plateau
until the accuracies of $E_{0}$ and $E_{1}$ are both sufficiently decreased,
after which the accuracies of all three energy levels collectively increase,
allowing the objective function to converge to its global minimum. While we
only illustrate this for one of the ten runs,
we observe that this qualitative behavior is typical for the other nine
runs. Besides the non-convexity reason from objective functions, other reasons
for this behavior could be related to two aspects of the weighted sum nature
of the SSVQE algorithm. First, the same ansatz circuit is applied to all
input states, meaning any change in the parameters has a direct impact on
all the expectation value terms simultaneously. Second, the energy levels
contribute unequally to the overall cost function, with the lower energy levels
contributing more than the higher energy levels, an effect that becomes more
pronounced due to the presence of the weight vector. Because qOMM enforces
orthogonality (at the global minimum) through its overlap terms, the energy
expectation value terms are associated with different independent parameter
values. This means that changes in parameters that directly affect one of
the energy expectation value terms in \eqref{eq:omm-obj} do not directly
change the others. They only change the overlap terms. This affords qOMM
a flexibility that allows it to spend much fewer iterations escaping from
the ``bad'' landscape regions. From Figure~\ref{fig:LiH-3states-zero}
and Figure~\ref{fig:LiH-2states-zero} we see that starting with a good
initial guess can reduce this effect, resulting in significantly improved
convergence.

We also briefly note that although we have chosen to measure convergence
speed in terms of the number of calls to the objective function, another
valid measure of convergence speed would be the total number of optimization
iterations if it can be assumed that one can parallelize multiple calls to the
objective function. This is because one optimization iteration can require
multiple calls to the objective function. In the noiseless results we have
presented here, each optimization iteration of the L-BFGS-B optimizer requires
$n$+1 calls to the objective function for $n$ total parameters. For problem
instances we have presented here where qOMM has more parameters than SSVQE,
the speedup of qOMM over SSVQE would be even greater by this measure. For
instance, in Figure~\ref{fig:LiH-3states-one-curve-random}, qOMM with 2-UCCSD has 144 total
parameters and SSVQE with 3-UCCSD has 72.

\begin{figure}[htb]
    \centering
    \includegraphics[width=\linewidth]{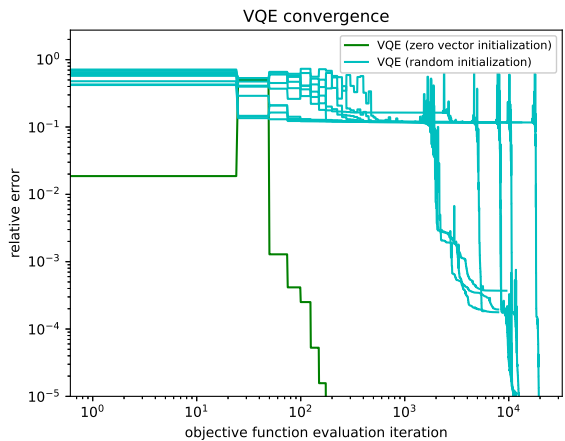}
    \caption{Convergence of  the relative error $\frac{\abs{E_{0,i} - E_{0,exact}}}{\abs{E_{0,exact}}}$ of VQE for the ground state problem of \ch{LiH}
    using a random parameter initialization}
    \label{fig:LiH-VQE}
\end{figure}

\emph{Initialization.} Initialization plays an important role in the
convergence for both qOMM and SSVQE. Even for the corresponding VQE ground
state problem with 1-repetition UCCSD, as in Figure~\ref{fig:LiH-VQE},
random parameter initialization on average converges in about 100 times
more objective function evaluations than the zero vector parameter
initialization. Both initializations in Figure~\ref{fig:LiH-VQE} converge
to a relative error $10^{-5}$. In excited state computing, the two
different initialization strategies for both qOMM and SSVQE not only
differ in the number of iterations but also differ in the convergence
results. Comparing Figures~\ref{fig:LiH-2states-zero} and \ref{fig:LiH-2states-one-curve-random}, for SSVQE with 2-repetition UCCSD, random parameter initialization converges more
than ten times slower than that of zero vector parameter initialization. A
similar result can be observed in Figures~\ref{fig:LiH-3states-zero} and \ref{fig:LiH-3states-one-curve-random} for SSVQE
with 3-repetition UCCSD. Besides the difference in the number of
iterations, optimization starting from random parameter initializations
sometimes cannot find the global minima, as can be seen from Table~\ref{tab:success_rate_table}. Overall, in all results included
in this paper, zero vector parameter initializations outperform their
random parameter initialization counterparts for both algorithms. However,
we still include numerical results for random parameter initializations to
demonstrate the different performance between qOMM and SSVQE. This is
because the zero vector initializations initialize the states to be
excitations from Hartree-Fock states and only work for a limited number of
low-lying states. If more excited states are needed, the zero vector
initialization could even be a bad initialization, which is due to the
fact that the excitation energies from Hartree-Fock may not be in the
consistent order with that of FCI excitation energies. This "zero vector"
initialization is demonstrated primarily not to advocate for its use as a
scalable initialization strategy, but to illustrate what the improved
convergence could look like if it can be assumed that one has a reasonable
initialization strategy. The flexibility of variational ansatzes in qOMM
would have another advantage when many excited states are needed. In qOMM,
we could use zero vector parameter initialization for a few states and
random parameter initialization for the remaining states without known good
initializations. Such a mixed initialization strategy could avoid being
trapped in bad local minima and benefit from the fast convergence of those
states with zero vector parameter initialization. SSVQE, unfortunately,
cannot use such a mixed initialization strategy due to the implicit
orthogonality among initialized states. This motivates further
investigation into developing and benchmarking initialization strategies
more sophisticated than the ones we consider in this work. For example, the MCVQE paper\cite{Parrish2019} proposes an efficient quantum circuit implementation of CIS states for excited state initializations. This initialization would likely improve the convergence results of the simulations for both qOMM and SSVQE presented in this paper. Furthermore, because the optimization stage of MCVQE can be seen as a special case of SSVQE that uses a CIS initialization and an equal weighting of the states, performing analogous simulations using a CIS initialization would offer insight into how the convergence of qOMM compares to that of MCVQE. The approach of using chemically-motivated initializations inspired from classical computational chemistry is likely to further improve the convergence of algorithms such as qOMM and SSVQE. The main challenge in this approach is that high level chemically-inspired initializations require efficient low-level circuit implementation, which is highly non-trivial.

\emph{Ansatz Circuit Depth.} 
Based on our numerical results, increasing
the ansatz circuit depth, \ie, increasing the repetition in UCCSD,
has two major impacts. First, the expressiveness increases as the
ansatz circuit depth increases. In the ground state VQE, as in
Figure~\ref{fig:LiH-VQE}, 1-repetition UCCSD is sufficient for the
ansatz to approximate the ground state to chemical accuracy. When
two states of \ch{LiH} are needed, 1-repetition UCCSD, as in
Figure~\ref{fig:LiH-one-curve-plots}, is not able to simultaneously
approximate the ground state and the first excited states,
whereas 2-repetition UCCSD is able to. From the qOMM results in
Figure~\ref{fig:LiH-2states-zero}, we know that 1-repetition UCCSD is able to approximate the ground state and the first excited state. When
three states are needed, as in Figure~\ref{fig:LiH-3states-zero}, we need
3-repetition UCCSD to simultaneously approximate three states. Again,
1-repetition UCCSD is still able to approximate any of these three
states. More generally, 2-UCCSD is sufficient for all \ch{LiH} cases studied when using qOMM, whereas SSVQE requires $n$-UCCSD in order to reliably find $n$ states. The results for the hydrogen square model in Figure~\ref{fig:H4-one-curve} demonstrates a similar pattern, except the situation is more severe for SSVQE, which requires more than $n$-UCCSD for finding $n$ states. A more detailed analytical description of the ability of the UCCSD ansatz
to simultaneously represent an increasing number of excited states
as the number of repetitions is increased is not yet known. qOMM
could adopt many UCCSDs, UCCSD with many repetitions, or a mixture of
them as its variational ansatz. SSVQE, however, could only benefit
from increasing the repetitions in UCCSD. The second impact of
ansatz circuit depth relates to optimization. Another field full of
non-convex optimization and parametrized ansatz is the deep neural
network. An interesting theoretical result~\cite{Allen-Zhu2019}
therein shows that enlarging the parameter space, \ie, increasing
the number of parameters, would make the optimization problem
easier, \ie, the energy landscape would be full of the approximated
global minimum. Although we have not established similar results
for quantum ansatz circuits, we observe the behavior from our
numerical results. In Figure~\ref{fig:LiH-3states-zero}, 1-repetition
UCCSD is able to characterize the three states for \ch{LiH} using
qOMM. However, all ten random parameter initializations got stuck as can be seen in Table~\ref{tab:success_rate_table}. As we increase the circuit
depth from 1-repetition to 2-repetition UCCSD, qOMM rapidly converges
to a global minimum for all ten random parameter initializations.

We conclude this section by discussing the circuit depth involved in
calculating the inner product terms in the qOMM objective function. Currently,
the only method we are aware of that involves running the circuit in
Figure~\ref{fig:inner_prod}. At a minimum, the inner product circuit will incur
a circuit depth twice that of the original ansatz. There will be additional
circuit depth due to the need to compile controlled versions of the ansatz to
the finite basis gate set of the machine. The extent of this additional depth
compared to the original ansatz will depend on a number of factors including
the basis gate set and qubit connectivity of the hardware, the choice of
ansatz, and the efficiency of the compilation algorithm used. We can give some
intuition for the circuit depths involved for the particular systems we studied here
for a particular basis gate set. Using the Qiskit compiler, the gate depth
for the 1-UCCSD ansatz used in our \ch{LiH} simulations is approximately 1400
gates when compiled to the basis gate set consisting of $R_z$, $\sqrt{X}$,
$X$, CNOT, and the identity. The corresponding circuit depth for computing
the inner product terms is approximately 19600 gates. These counts would
be scaled roughly by a factor of $n$ for $n$-UCCSD. This underscores the
importance of future work that addresses how to reduce these gate counts. In
particular, finding (if they exist) more efficient implementations of the
inner product computations will be important for the applicability of qOMM
in the NISQ era. We emphasize that while there are several known subroutines for computing the overlap of two states for other applications such as quantum machine learning\cite{Cincio_2018}, these generally compute real quantities of the form $\abs{\bra{\psi_1}\ket{\psi_2}}$, whereas for qOMM we need the full complex quantity $\bra{\psi_1}\ket{\psi_2}$. The observation that SSVQE appears to require an increasingly
expressive ansatz for increasingly larger systems and an increasing number
of states is also a concern. Future work that addresses how to mitigate this
effect will be important for the applicability of SSVQE in the NISQ era. In particular, it has been shown that certain conditions can induce cost function landscapes with gradient magnitudes that vanish exponentially (\emph{i.e.} barren plateaus) with the number of qubits, rendering convergence infeasible. This has been shown to occur in instances where one uses a hardware-efficient ansatz in combination with a non-local cost function\cite{Cerezo2021} and in the presence of noise when one uses an ansatz with circuit depth that scales linearly or faster with the number of qubits.\cite{Wang2021}. Because both of these types of barren plateaus worsen with increasing circuit depth, the extent to which the two methods studied in this paper are susceptible to this effect will likely depend on how the circuit depth required for each algorithm to converge scales with increasing problem size. Such an investigation would require the simulation of larger chemical systems and would be an interesting direction of future research. We also note that there are various strategies that could be employed to facilitate the study of larger systems by representing the active space with as few qubits as possible, effectively delaying the onset of barren plateaus. A simple, well-known example of this idea was used in this work when we froze the two core orbitals in \ch{LiH} to reduce the problem size from 12 qubits to 10. Another strategy would be to optimize the basis set under a fixed qubit budget, as is done in OptOrbFCI in the context of classical computational chemistry\cite{doi:10.1021/acs.jctc.0c00613}. Generalizing this work to quantum variational algorithms such as VQE, SSVQE, and qOMM will be the topic of future work.

\section{Conclusion}
\label{sec:conclusion}

In this work we have proposed a method which can calculate the low-lying
eigenvalue/eigenstate pairs of a Hermitian operator on quantum hardware.
We have classically simulated the algorithm for three different electronic
structure Hamiltonians: \ch{H2}, \ch{LiH}, and the hydrogen square lattice. We
have compared it to SSVQE, another algorithm with the same goal, but
one which enforces the orthogonality of the input states explicitly at
every optimization step.  We showed that a small problem such as \ch{H2}
is not difficult enough to meaningfully distinguish any differences in
performance for these two approaches when using a random initialization
strategy, but when one compares them to a more moderately-sized problems
such as 10 qubit \ch{LiH}, noticeable
differences in performance begin to emerge. In this scenerio, qOMM converges
with fewer objective function evaluations and with a less expressive ansatz in all but one of the simulations
considered here. Each of these methods
incurs additional circuit depth beyond the ground state problem that should
be addressed by future works in order to increase their applicability in the
NISQ era. This corresponds to finding more efficient methods for computing
the inner product terms in qOMM and reducing the ansatz circuit depth needed
for SSVQE to converge efficiently. One potential direction for attempting
to improve SSVQE would be to investigate the extent to which incorporating
adaptive ansatz strategies~\cite{Grimsley2019, Tang2021, Yordanov2021}
would alleviate this apparent feature of SSVQE. These methods were developed
in order to find depth-efficient ansatz circuits for problem instances for
which doing so by manual heuristic guesswork is difficult. This is precisely
the problem from which we observed SSVQE to suffer in this work. qOMM could
benefit from these methods as well. Additionally, one could investigate
whether or not modifying the SSVQE weight vector to adaptively change over
the course of the optimization could mitigate the objective function plateau
effect we observed in our simulations.

The \ch{LiH} simulations demonstrate that in the context of
these types of excited states methods, even a crude initialization
strategy can greatly reduce not only the number of optimization iterations
needed to converge, but also the circuit depth needed to do so. Another
potentially interesting topic would be to develop and benchmark more powerful and versatile initialization strategies that are taylored specifically to these excited states methods. This would also improve the applicability of these methods in the NISQ era.

%% classes. This should be given within the "acknowledgement"
 
\begin{acknowledgement}
    The work is supported in part by the US National Science Foundation
    under award CHE-2037263 and by the US Department of
    Energy via grant DE-SC0019449.
\end{acknowledgement}

% \begin{suppinfo}

% A listing of the contents of each file supplied as Supporting Information
% should be included. For instructions on what should be included in the
% Supporting Information as well as how to prepare this material for
% publications, refer to the journal's Instructions for Authors.

% The following files are available free of charge.

% %\begin{itemize}
% %    \item Filename: brief description
% %    \item Filename: brief description
% %\end{itemize}

% \end{suppinfo}


\bibliography{reference}

%%% Add Appendix
%\newpage
\appendix

\section{Additional Convergence Plots}\label{appendix: Convergence Plots}

Here we provide all ten runs for each of the randomly initialized tests for which only one run was shown in Section~\ref{sec:num}. We also provide some additional tests such as \ch{H_{2}} and the hydrogen square model at stretched bond distances and up to 7 states of \ch{LiH}. The success rates for the noise-free simulations are summarized in Table~\ref{tab:success_rate_table} in Section~\ref{sec:num}.

\begin{figure*}[htp]
    \centering
    \begin{subfigure}{0.49\linewidth}
        \centering
        \includegraphics[width=\linewidth]{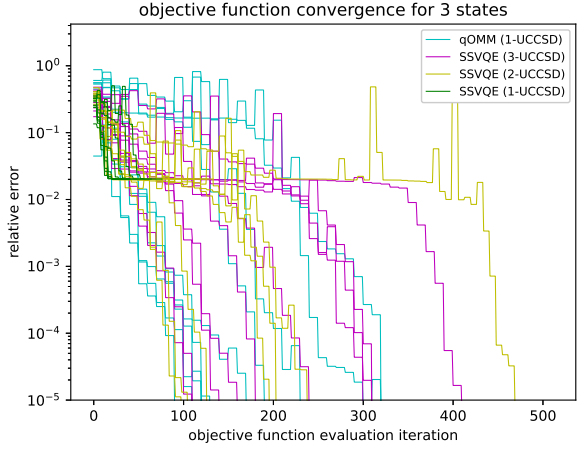}
        \caption{0.735 {\AA} bond distance}
        \label{fig:H2-3states-random}
    \end{subfigure}
    \begin{subfigure}{0.49\linewidth}
        \centering
        \includegraphics[width=\linewidth]{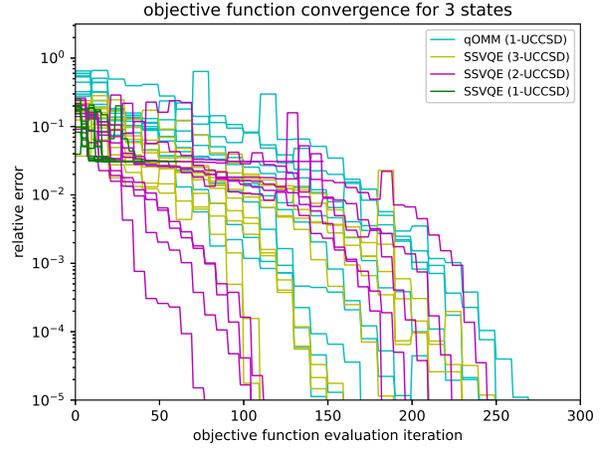}
        \caption{1.47 {\AA} bond distance}
        \label{fig:H2-1.47-3states-random}
    \end{subfigure}
    \caption{Convergence of the relative error $\frac{\abs{f_{i} - f_{exact}}}{\abs{f_{exact}}}$ of qOMM and SSVQE for \ch{H_2} using a random parameter initialization.}
    \label{fig:H2}
\end{figure*}

\begin{figure*}[htb]
    \centering
    \begin{subfigure}{0.49\linewidth}
        \centering
        \includegraphics[width=\linewidth]{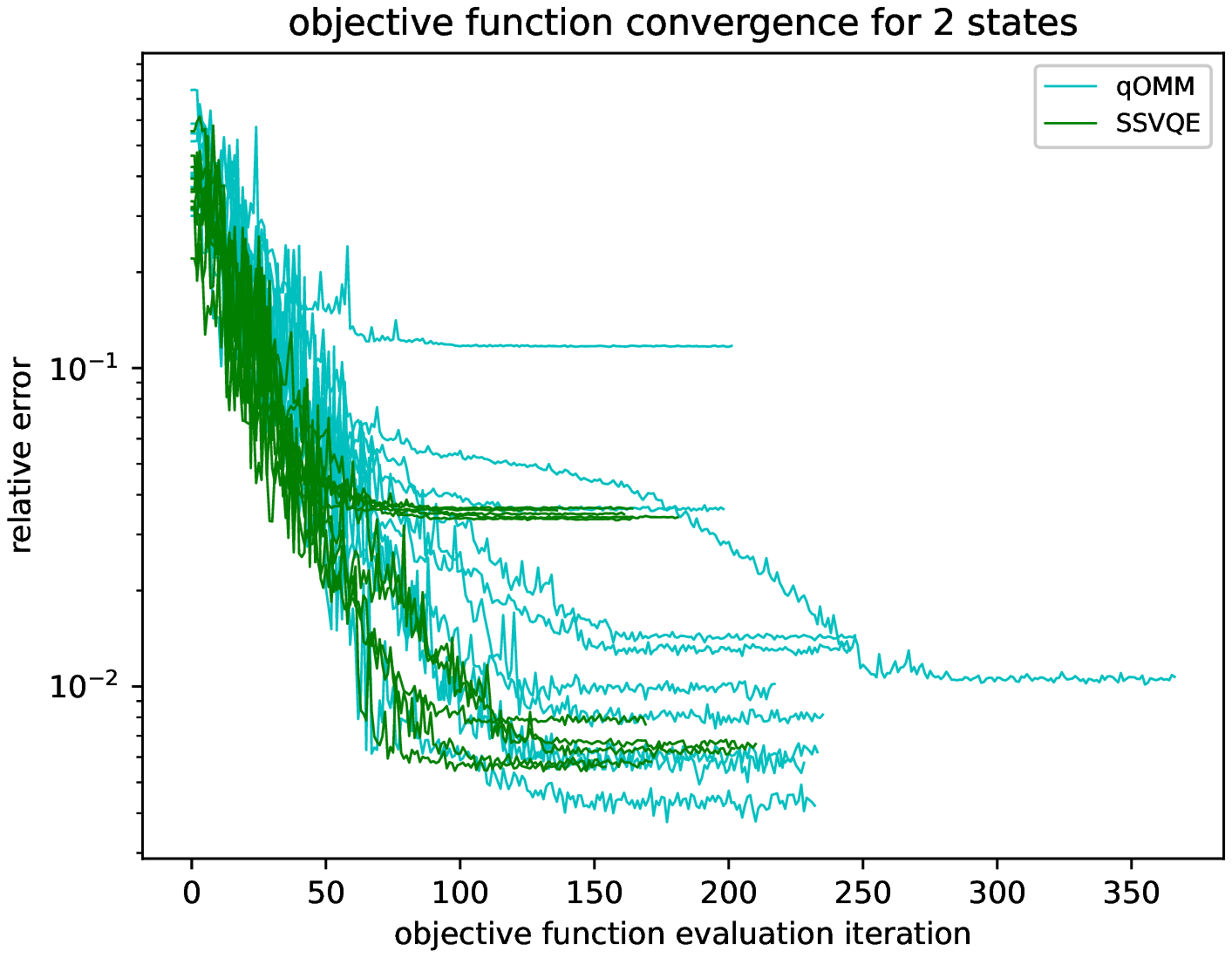}
        \caption{Two states}
        \label{fig:H2-2states-noisy}
    \end{subfigure}
    \begin{subfigure}{0.49\linewidth}
        \centering
        \includegraphics[width=\linewidth]{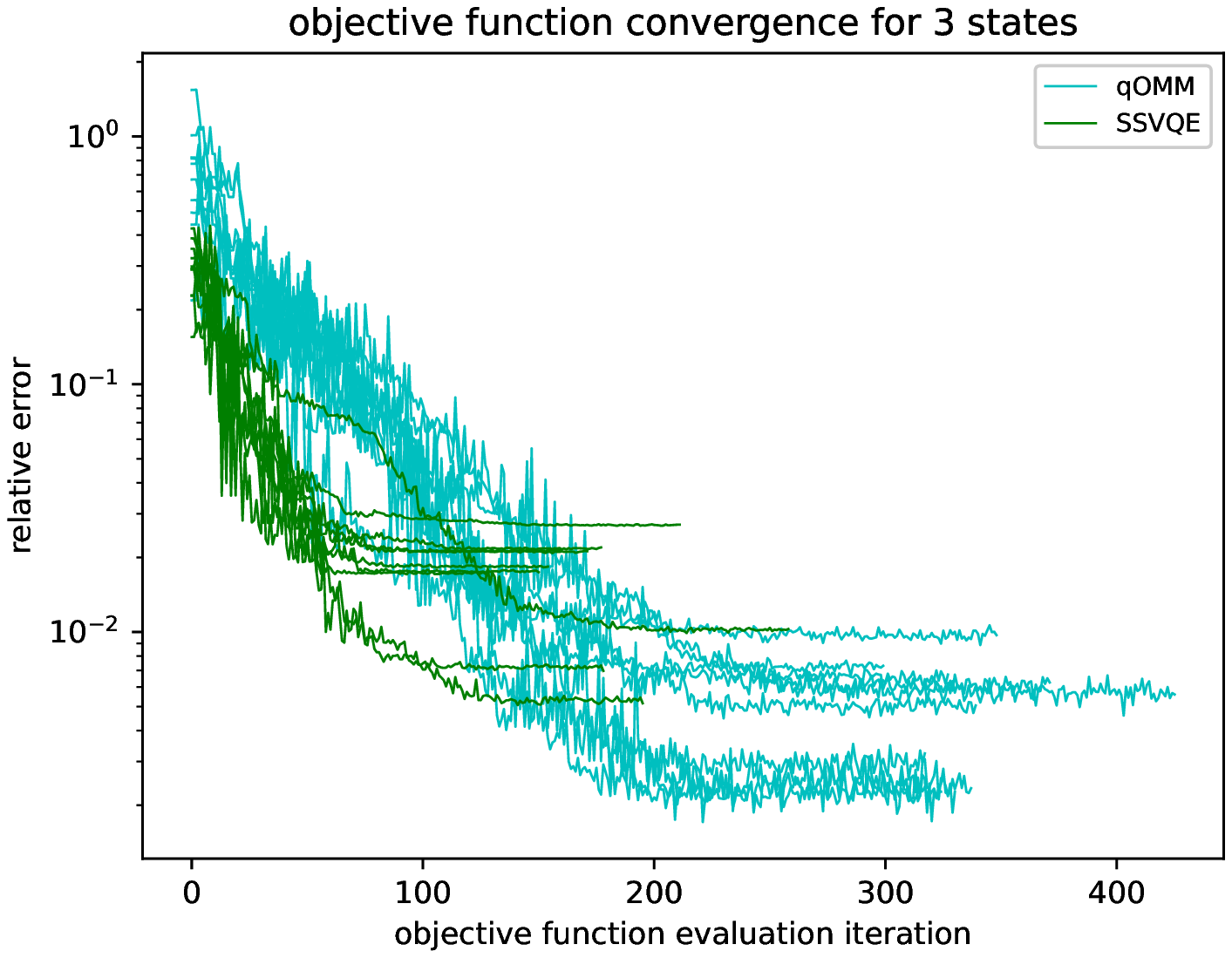}
        \caption{Three states}
        \label{fig:H2-3states-noisy}
    \end{subfigure}
    \caption{Convergence of  the relative error $\frac{\abs{f_{i} - f_{exact}}}{\abs{f_{exact}}}$ of qOMM and SSVQE for \ch{H_2} at an interatomic distance of 0.735 {\AA}, where each circuit gate is modelled as having a probability of local depolarizing error $p_{error} = 0.001$.}
    \label{fig:H2-noisy}
\end{figure*}

\begin{figure*}[htp!]
    \centering
    \begin{subfigure}{0.49\linewidth}
        \centering
        \includegraphics[width=\linewidth]{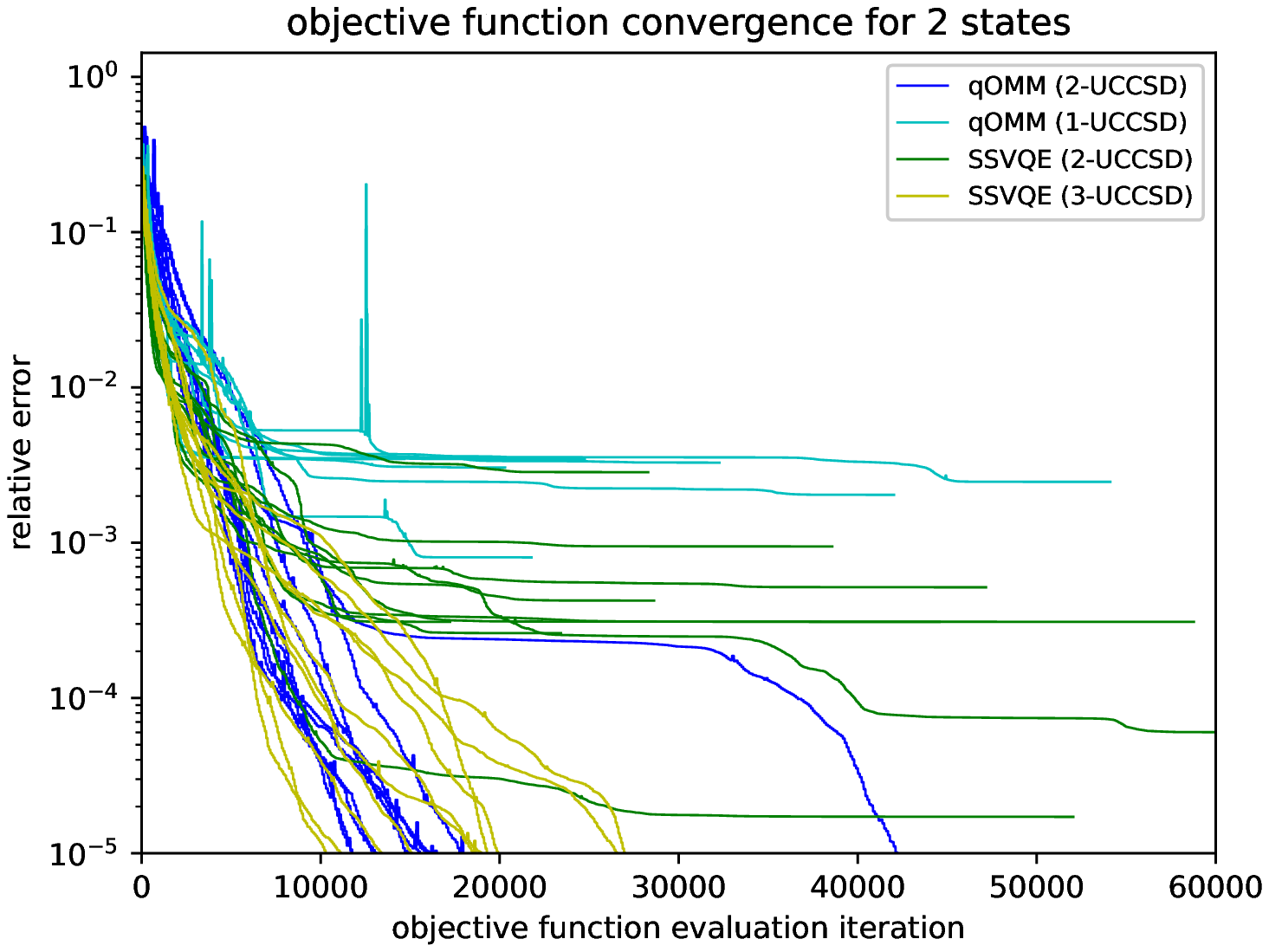}
        \caption{Two states}
        \label{fig:H4-2states}
    \end{subfigure}
    \begin{subfigure}{0.49\linewidth}
        \centering
        \includegraphics[width=\linewidth]{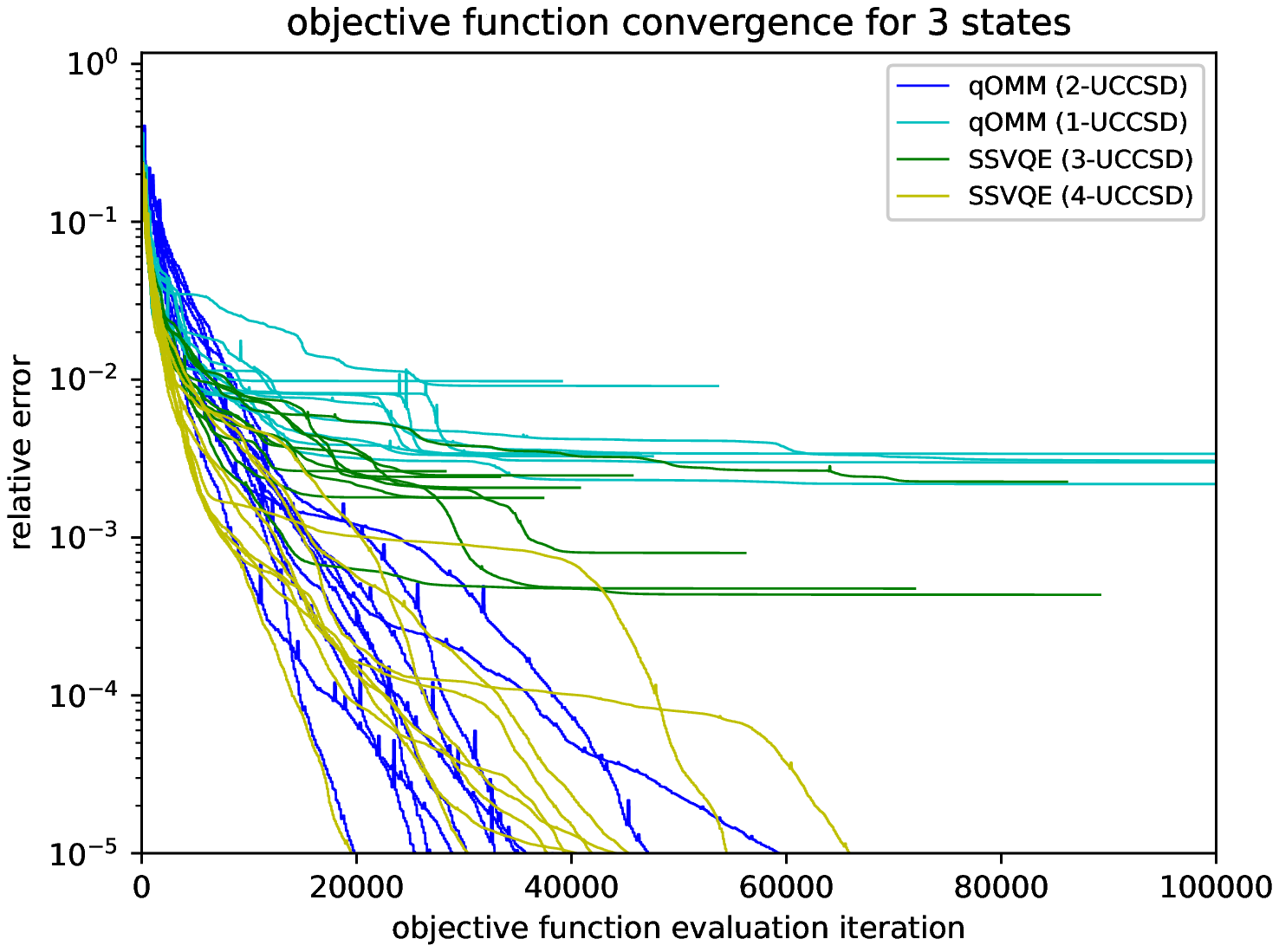}
        \caption{Three states}
        \label{fig:H4-3states}
    \end{subfigure}
    \caption{Convergence of the relative error $\frac{\abs{f_{i} - f_{exact}}}{\abs{f_{exact}}}$ of qOMM and SSVQE for the hydrogen square model at an interatomic distance of 1.23 {\AA} using a random parameter initialization.}
    \label{fig:H4}
\end{figure*}

\begin{figure*}[htp!]
    \centering
    \begin{subfigure}{0.49\linewidth}
        \centering
        \includegraphics[width=\linewidth]{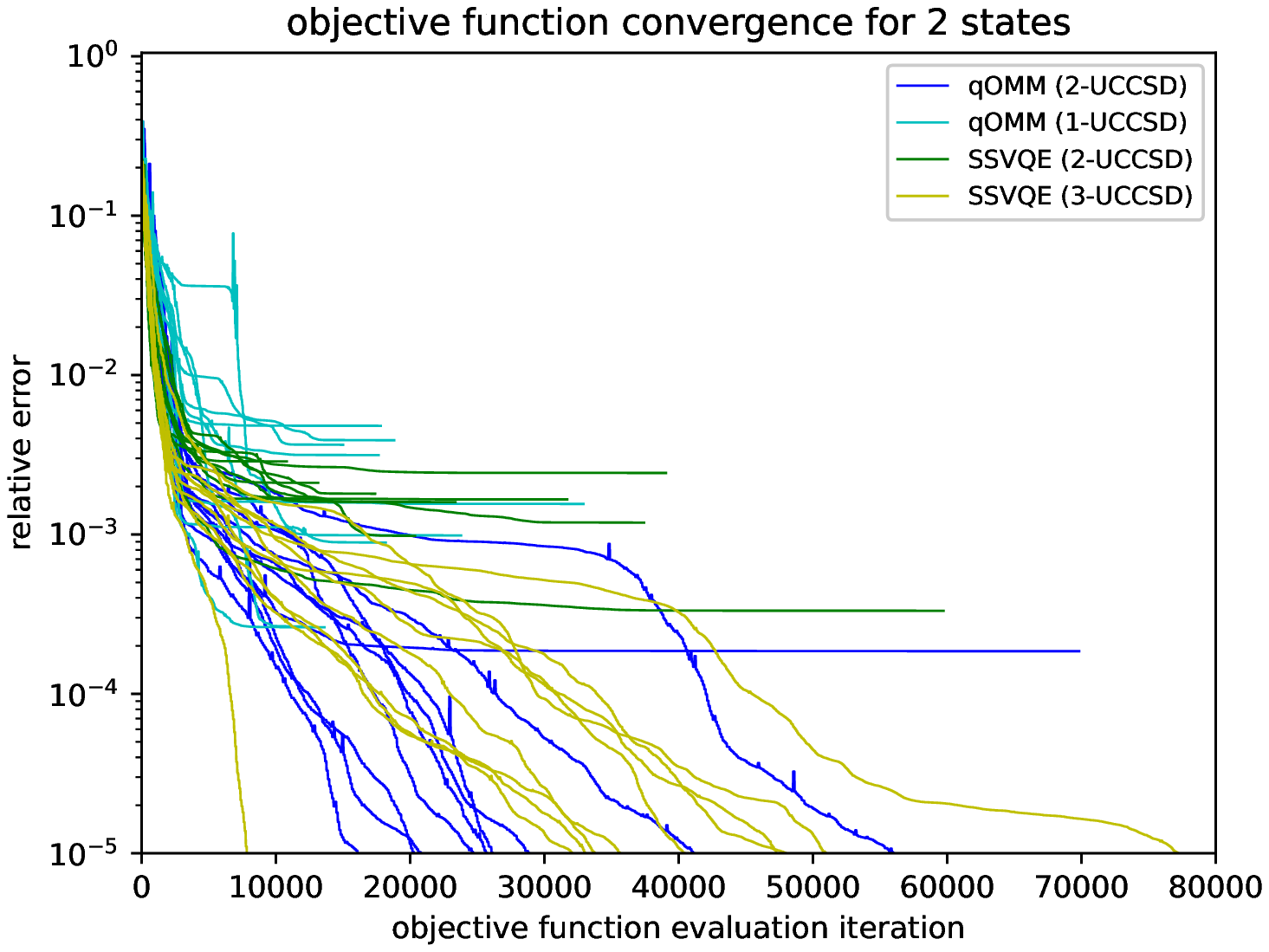}
        \caption{Two states}
        \label{fig:H4-2states-stretched}
    \end{subfigure}
    \begin{subfigure}{0.49\linewidth}
        \centering
        \includegraphics[width=\linewidth]{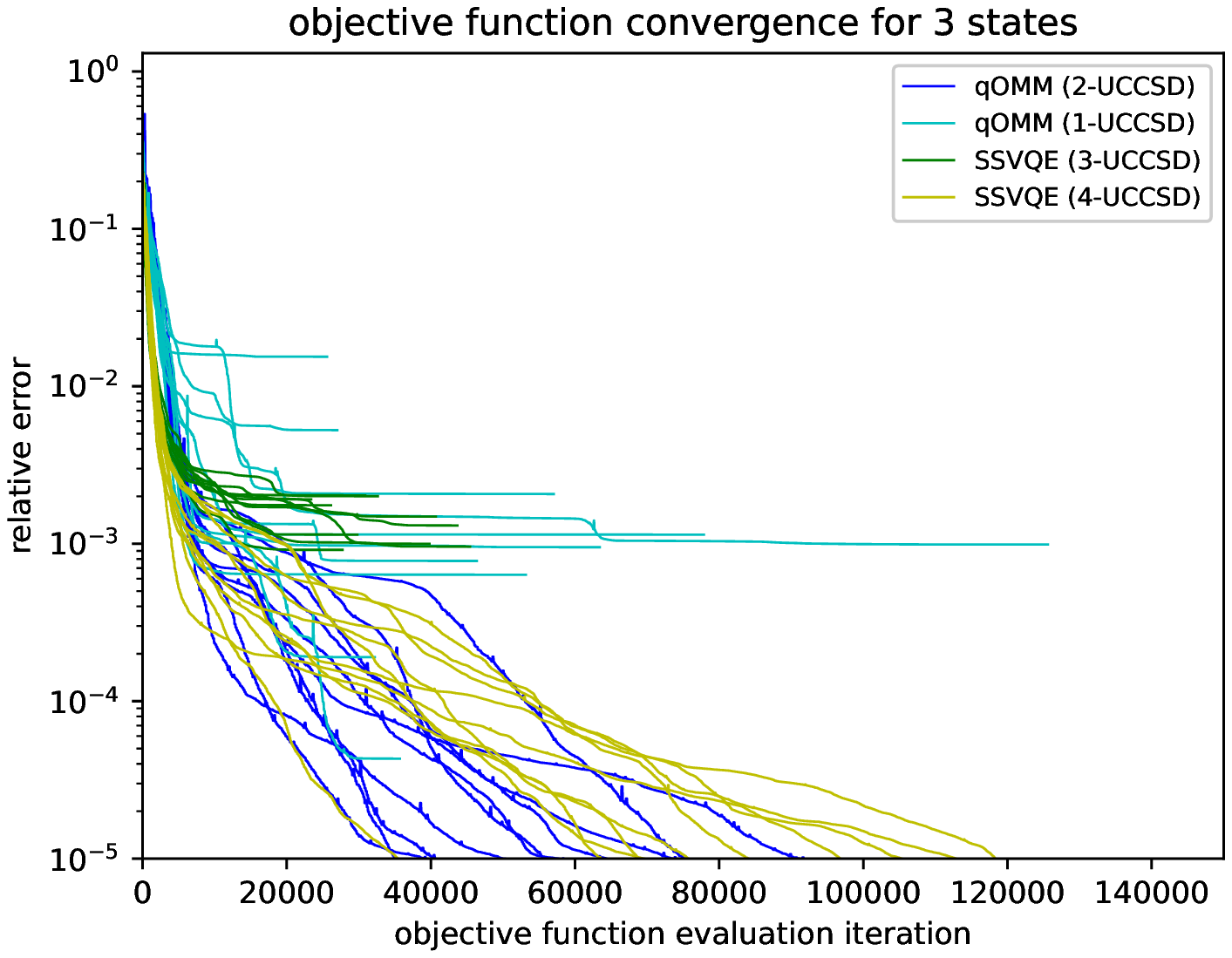}
        \caption{Three states}
        \label{fig:H4-3states-stretched}
    \end{subfigure}
    \caption{Convergence of the relative error $\frac{\abs{f_{i} - f_{exact}}}{\abs{f_{exact}}}$ of qOMM and SSVQE for the hydrogen square model at an interatomic distance of 2.46 {\AA} using a random parameter initialization.}
    \label{fig:H4-stretched}
\end{figure*}

\begin{figure*}[htb]
    \centering
    \begin{subfigure}{0.49\linewidth}
        \centering
        \includegraphics[width=\linewidth]{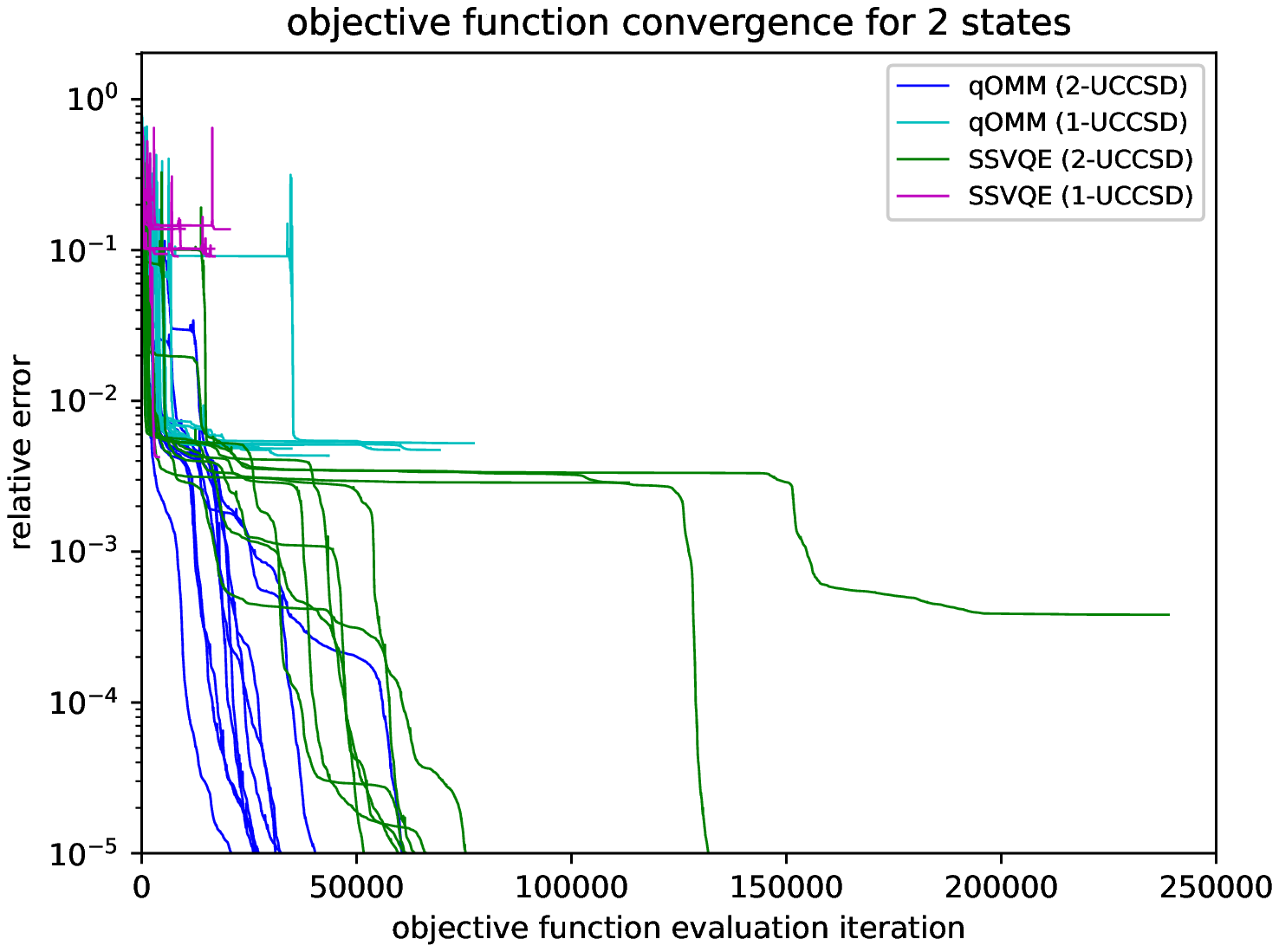}
        \caption{Two states}
        \label{fig:LiH-2states}
    \end{subfigure}
    \begin{subfigure}{0.49\linewidth}
        \centering
        \includegraphics[width=\linewidth]{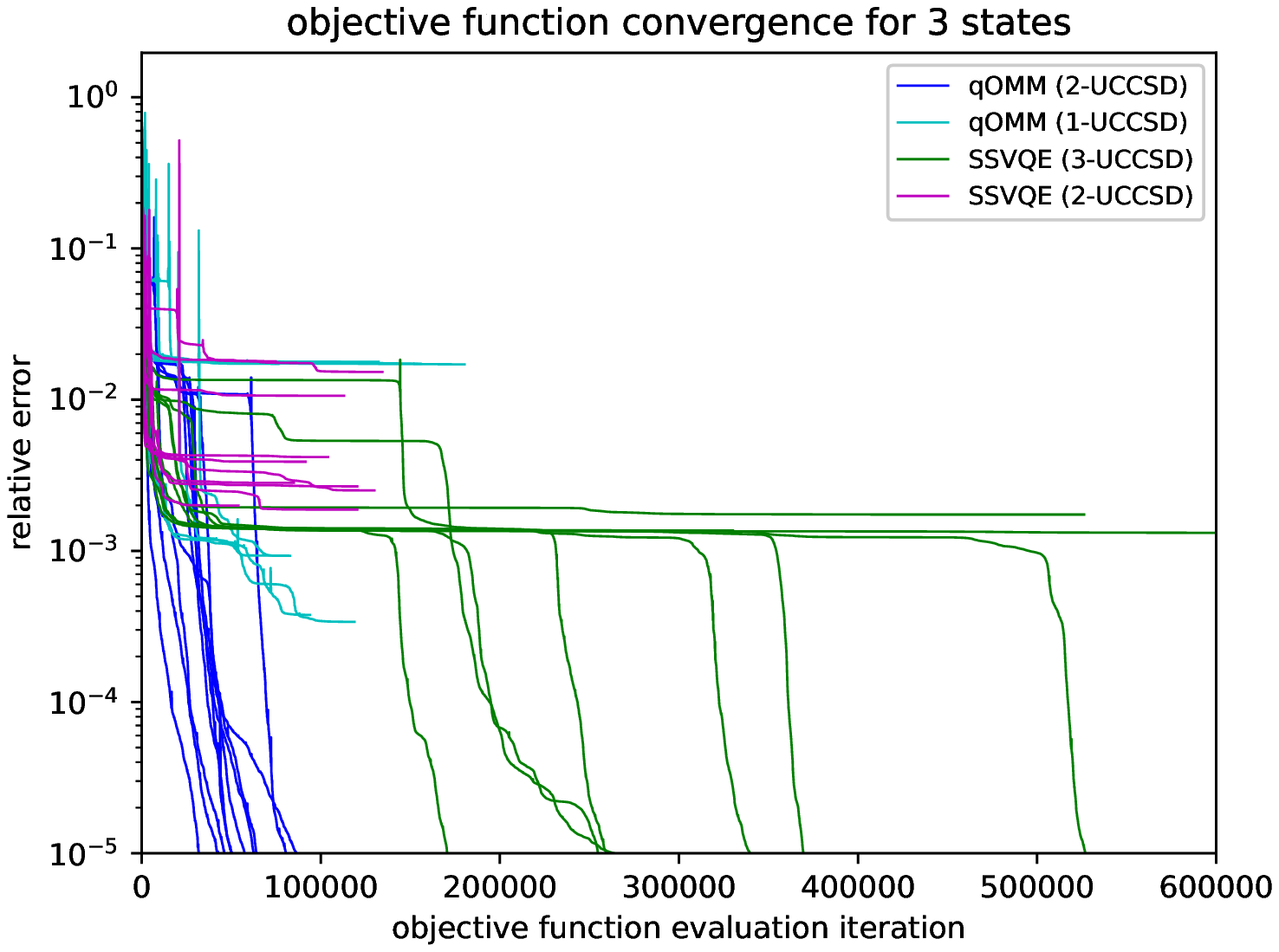}
        \caption{Three states}
        \label{fig:LiH-3states}
    \end{subfigure}
        \centering
    \begin{subfigure}{0.49\linewidth}
        \centering
        \includegraphics[width=\linewidth]{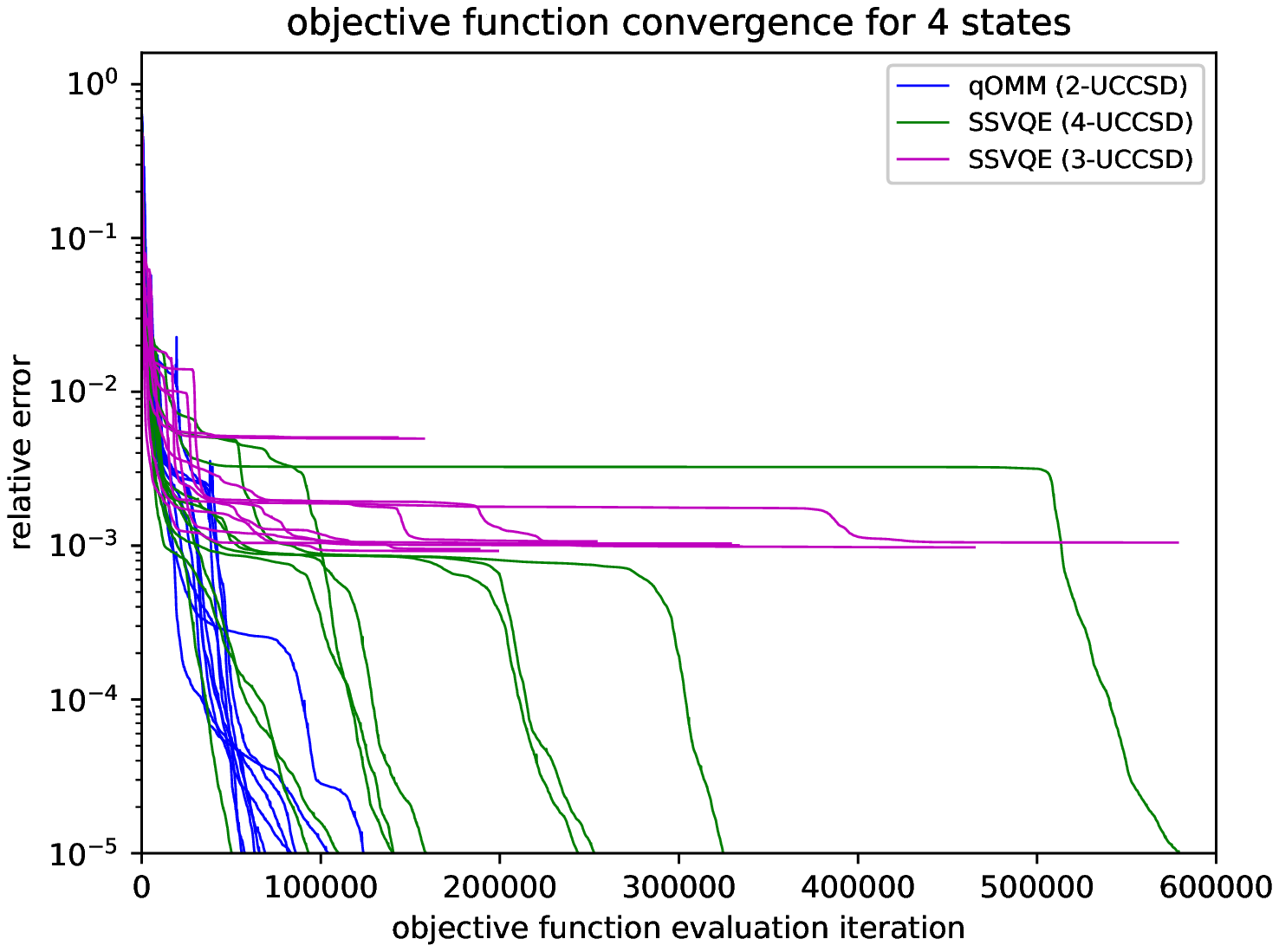}
        \caption{Four states}
        \label{fig:LiH-4states}
    \end{subfigure}
    \begin{subfigure}{0.49\linewidth}
        \centering
        \includegraphics[width=\linewidth]{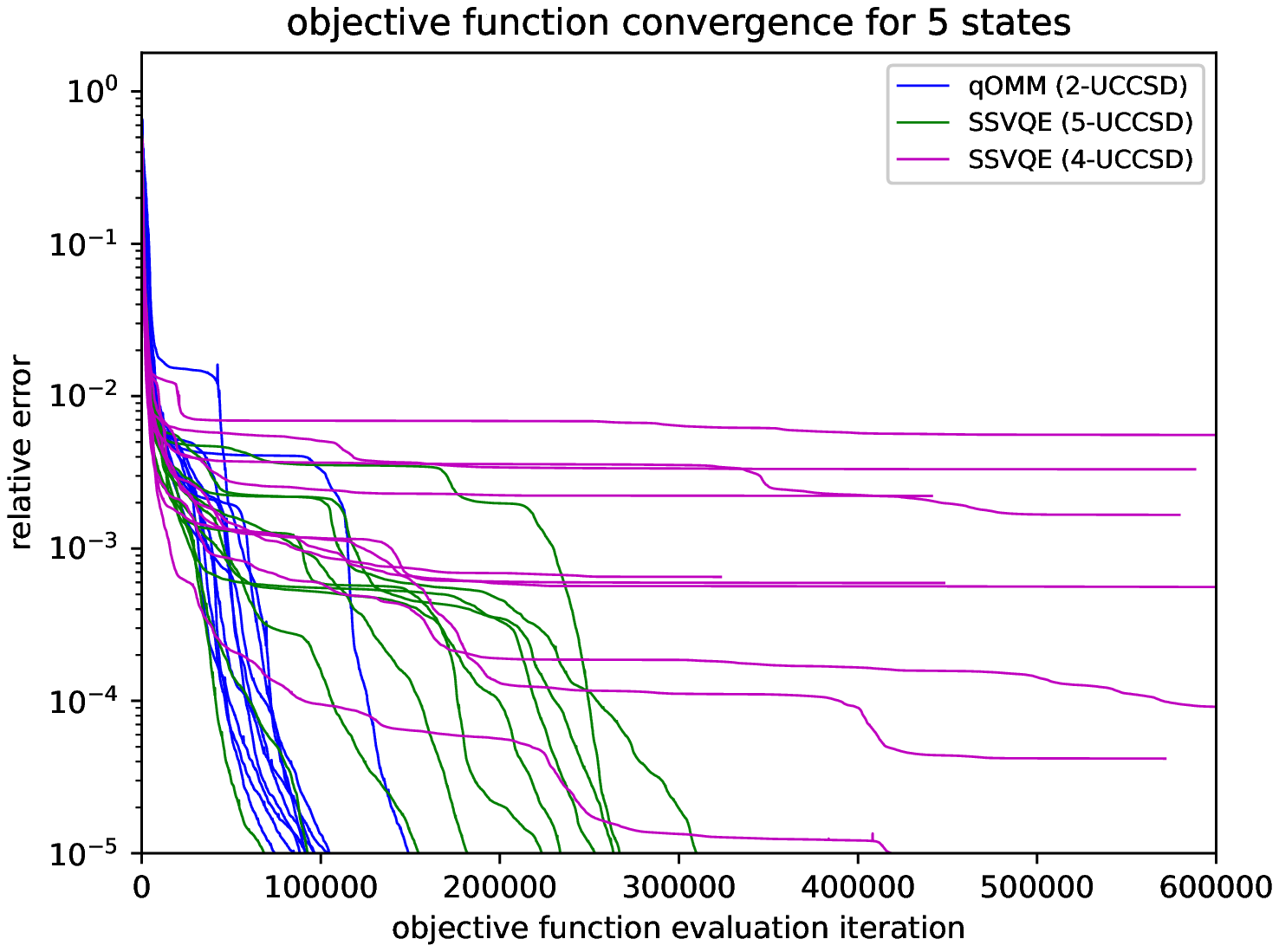}
        \caption{Five states}
        \label{fig:LiH-5states}
    \end{subfigure}
        \centering
    \begin{subfigure}{0.49\linewidth}
        \centering
        \includegraphics[width=\linewidth]{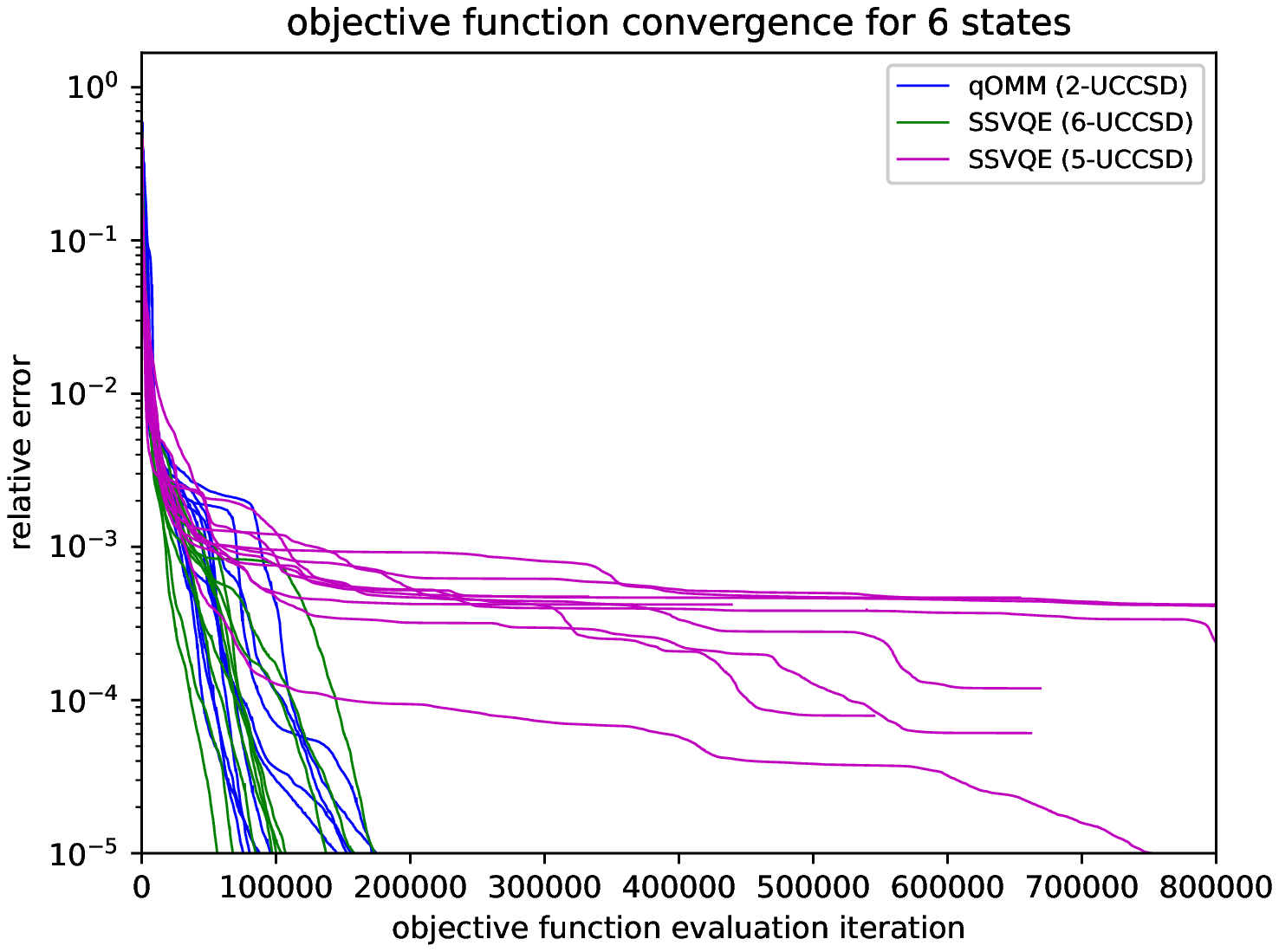}
        \caption{Six states}
        \label{fig:LiH-6states}
    \end{subfigure}
    \begin{subfigure}{0.49\linewidth}
        \centering
        \includegraphics[width=\linewidth]{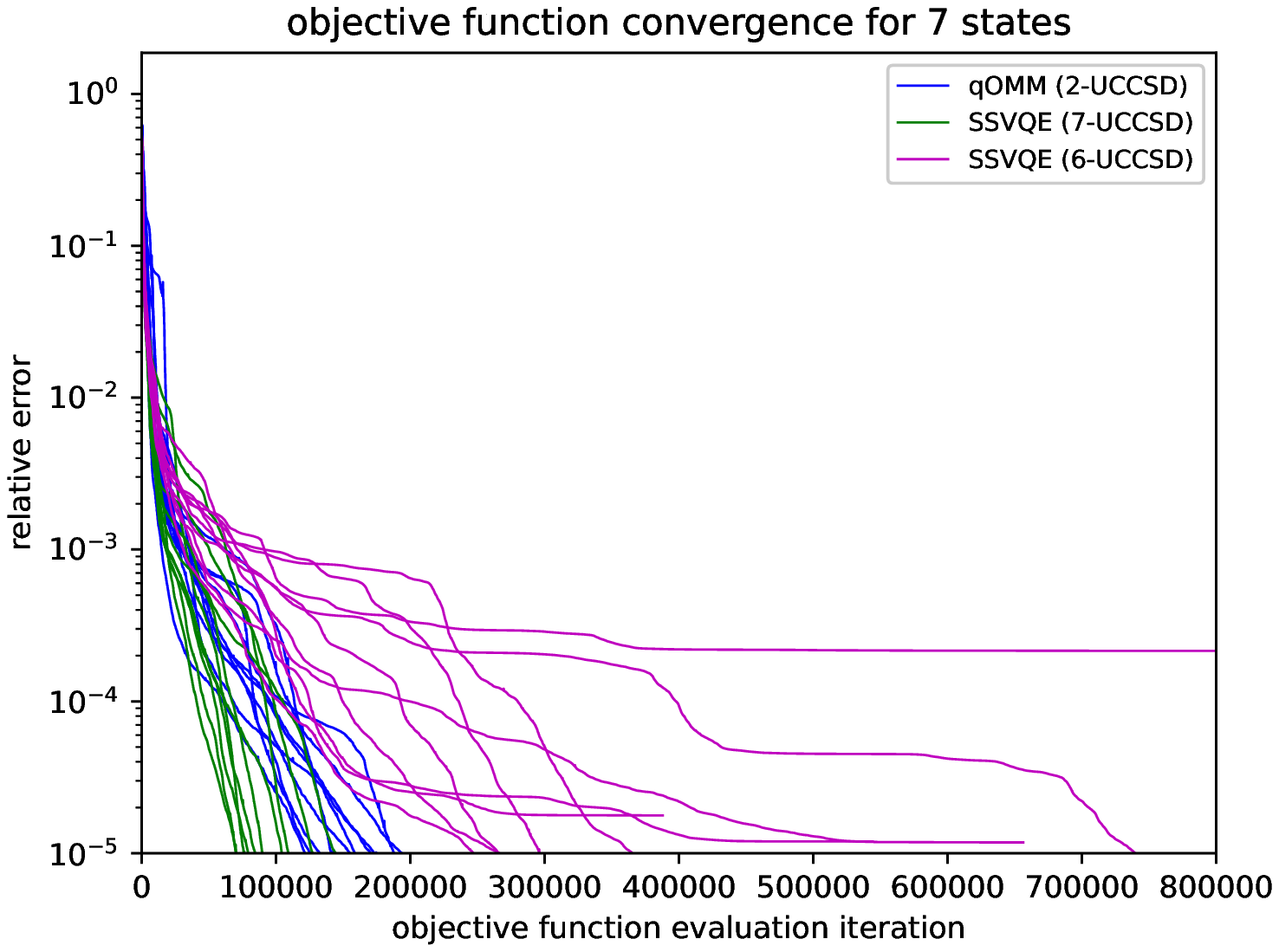}
        \caption{Seven states}
        \label{fig:LiH-7states}
    \end{subfigure}
    \caption{Convergence of the relative error $\frac{\abs{f_{i} - f_{exact}}}{\abs{f_{exact}}}$ of qOMM and SSVQE for \ch{LiH} using randomly initialized parameters.}
    \label{fig:LiH-random-extended}
\end{figure*}

\end{document}